\documentclass[preprint,preprintnumbers,tightenlines,amsmath,amssymb]{revtex4}
\usepackage{graphicx}% Include figure files
\usepackage{dcolumn}% Align table columns on decimal point
\usepackage{bm}% bold math
\usepackage{cancel}
\begin{document}
%\preprint{APS/123-QED}

\title{Reduction of one-loop $n$-point integrals}% Force line breaks with \\

\author{Kwangwoo Park}
\email{kwpark@mail.smu.edu}
\affiliation{Department of Physics, Southern Methodist University, Dallas, TX, 75275-0175, USA}
\date{\today}% It is always \today, today,

\begin{abstract}
In this paper, we focus on the analytical expressions of both three--point and four--point integrals for the case of a small Gram determinant. We also investigate the numerical efficiency of $n$-point integrals for $n\ge5$. 
One--loop five--point and higher--point integrals are crucial for accurate predictions of LHC physics, such as Higgs searches involving background process with three or more final states. The expressions of five--point and six--point integrals have been calculated in Ref.~\cite{Denner:2005nn}
using the fact that the loop momentum can be expressed as a linear combination of the external momenta.
We first present a proof of the results of Ref.~\cite{Denner:2005nn} using a new reduction formula. NLO calculations typically suffer numerical instabilities in specific phase space regions when individual kinematic variables are small; this leads to a vanishing Gram determinant. In these problematic regions, we introduce a technique that resolves the numerical instability analytically instead of using a numerical iteration. Explicit expressions of three--point and four--point integrals in the limit of small Gram determinants are provided by this new method. Furthermore, we demonstrate that one--loop $n$--point integral (with $n\ge6$) can always be reduced to six $(n-1)$--point integrals in general. This dramatically reduces the CPU time for numerical computations compared to other methods. Additionally, the numerical uncertainty originating from the computation of the higher dimensional Cayley matrix can be removed. We present general reduction formulas for five--point and higher--point scalar, vector, and tensor integrals at the one--loop level.
\end{abstract}

\maketitle
\def\.{ \! \cdot \! }
\def\ee{{{\bf e}_{0}}}
\def\ex{{{\bf e}_{1}}}
\def\ey{{{\bf e}_{2}}}
\def\ez{{{\bf e}_{3}}}
\def\A{\mathcal{A}}
\def\N{\mathbb{N}}
\def\M{\mathbb{M}}
\def\F{\mathcal{F}}
\def\P{\mathcal{P}}
\setlength{\unitlength}{1mm}

\section{\label{sec1}Introduction}
Recent experiments have produced precise predictions for Standard Model (SM) observables in both the electroweak (EW) and QCD sectors. Many of these observables involve multi-particle final-states. To make accurate NLO predictions for these observable, it is necessary to compute loop diagrams with many legs. The key ingredient of these calculations is the $n$--point loop integral. Accurate theoretical predictions therefore rely on accurate and efficient evaluation of the various $n$--point integrals which we encounter.

In this paper, we study one-loop integrals that are essential for the NLO radiative corrections. We are particularly interested in one--loop integrals for processes with three or more final-state particles. The various $n$--point integrals necessary have been calculated partially or completely~\cite{Davydychev:1991va,Denner:1991kt,Bern:1993kr,Binoth:1999sp,Denner:2002ii}. If the conventional Passarino--Veltman reduction~\cite{Passarino:1978jh} scheme is used, we must address the difficulty of a vanishing Gram determinant. Also, if we are computing heavy quarks, our results must be solved for general mass values, and this drastically complicates the kinematics.

This paper is organized as follows: In Sec.~\ref{sec2}, we prove the result of the Ref.~\cite{Denner:2005nn} for $n$--point integrals ($n\ge 5$) with algebraic formalism. In Sec.~\ref{sec3}, we drive the explicit expressions of three--point and four--point integrals for the case of a vanishing Gram determinant. In Sec.~\ref{sec4}, the expressions of $n$-point integrals with $n\ge5$ will be provided. In Sec.~\ref{sec4}, the numerical improvement will be summarized.

\section{\label{sec2}Formalism }
\subsection{\label{sec2-1}For the case of linear dependence on external momenta}
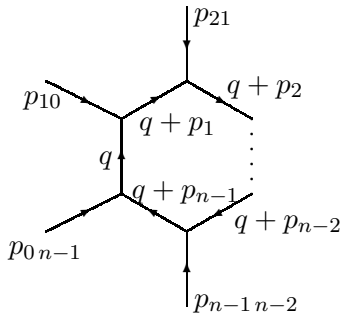
\begin{figure}
\begin{center}
\begin{picture}(60,50)
\qbezier(30,10)(30,10)(38.66,15)
\qbezier(30,10)(30,10)(21.34,15)
\put(21.34,15){\line(0,1){10}}
%\put(38.66,15){\line(0,1){10}}
\qbezier(30,30)(30,30)(21.34,25)
\qbezier(30,30)(30,30)(38.66,25)
\put(30,0){\vector(0,1){6}}
\put(30,6){\line(0,1){4}}
\qbezier(11.34,10)(11.34,10)(21.34,15)
\put(11.34,10){\vector(2,1){6}}
\qbezier(11.34,30)(11.34,30)(21.34,25)
\put(11.34,30){\vector(2,-1){6}}
\put(30,40){\vector(0,-1){6}}
\put(30,34){\line(0,-1){4}}
%\qbezier(48.66,30)(48.66,30)(38.66,25)
%\put(48.66,30){\vector(-2,-1){6}}
\put(38.66,25){\circle*{0.5}}
\put(38.66,23){\circle*{0.5}}
\put(38.66,21){\circle*{0.5}}
\put(38.66,19){\circle*{0.5}}
\put(38.66,17){\circle*{0.5}}
\put(38.66,15){\circle*{0.5}}
\put(25.67,12.5){\vector(-2,1){1}}
\put(21.34,20){\vector(0,1){1}}
\put(25.67,27.5){\vector(2,1){1}}
\put(34.33,27.5){\vector(2,-1){1}}
\put(34.33,12.5){\vector(-2,-1){1}}
\put(8.34,27){$p_{10}$}
\put(6.34,7){$p_{0\,n-1}$}
\put(31,0){$p_{n-1\,n-2}$}
\put(36.33,10.5){$q+p_{n-2}$}
\put(35.33,28.5){$q+p_2$}
\put(31,38){$p_{21}$}
\put(18.34,19){$q$}
\put(23.67,23.5){$q+p_{1}$}
\put(22.67,14.5){$q+p_{n-1}$}
\end{picture}
\end{center}
\caption{General one-loop $n$-point diagram for the standard form of $T^{n}_{\mu \nu \cdots}$ with $p_{ij}=p_i-p_j$. The arrows show the direction of momentum rather than the particle's current.}\label{Nloopfig}
\end{figure}

We begin with some standard definitions for the one--loop scalar $n$--point integral following Ref.~\cite{Denner:2005nn}. The $n$-point scalar integral in standard form is:
\begin{align}\label{st}
T^n_0 =\frac{(2 \pi \mu)^{4-D}}{i\pi^2} \int d^D q \frac{1}{N_0 N_1 \cdots N_{n-1}} \quad,
\end{align}
with the denominator factors:
\begin{align}\label{st0}
N_k=(q+p_k)^2 - m_k^2 +i \epsilon \quad, \qquad k=0,\cdots,n-1\quad , \qquad p_0=0 \quad.
\end{align}
In general, we also use alphabetic notation for the $n$--point functions; $T^1=A$, $T^2=B$, $T^3=C$, $T^4=D$, and so on.
For the reduction formula, the integral with one denominator factor omitted is very useful; $T^{n-1}_0 (k)$
denotes that $k^{\rm th}$ denominator $N_k$ is omitted in Eq.~\eqref{st}:
\begin{align}\label{T(k)}
T^{n-1}_0 (k)=\frac{(2 \pi \mu)^{4-D}}{i\pi^2} \int d^D q \frac{1}{N_0 \cdots N_{k-1} N_{k+1} \cdots N_{n-1}}\quad.
\end{align}
As shown in Ref~\cite{Denner:2005nn}, it is assumed that $T^n_0$ can be spanned by $T^{n-1}_0(k)$, which we will prove with a different method:
\begin{align}\label{span0}
T^n_0= \sum_{i=0}^{n-1} a_i \, T^{n-1}_0(i)\quad.
\end{align}
If we find all $a_i$ coefficients, then we can reduce any $n$-point scalar integral to a number of $(n-1)$--point scalar integrals.

Let us start from the integrand of Eq.~\eqref{st} with unknown $a_i$ coefficients:
\begin{align}\label{ids}
\frac{1}{N_0\cdots N_{n-1}}=\frac{a_0}{N_1\cdots N_{n-1}}+\frac{a_1}{N_0N_2 \cdots N_{n-1}}+\cdots+\frac{a_n}{N_0\cdots N_{n-2}}=\frac{\sum_{i=0}^{n-1} a_i N_i }{N_0\cdots N_{n-1}} \quad .
\end{align}
This must be true for arbitrary $q^\mu$. Comparison of numerators in both sides gives:
\begin{align}\label{cond1}
1=\sum_{i=0}^{n-1} a_i N_i&=\sum_{i=0}^{n-1}a_i\left( q^2-2 q_\mu p_i^\mu + p_i^2-m_i^2\right)\nonumber \\
&=q^2 \left(\sum_{i=0}^{n-1}a_i\right) -2 q_\mu \kappa^\mu+\sum_{i=0}^{n-1}(p_i^2-m_i^2) a_i \quad,
\end{align}\label{kappa}
where $\kappa^\mu$ is:
\begin{align}
\kappa^\mu=\sum_{i=0}^{n-1} a_i p_i^\mu \quad.
\end{align}
In order for Eq.~\eqref{cond1} to always be true for all $q^\mu$, the coefficients of $q^\mu$ and $q^2$ must vanish, while the constant term should be equal to $1$. This implies:
\begin{align}
\sum_{i=0}^{n-1}a_i=0 \quad, \label{cons-1} \\
\sum_{i=0}^{n-1} (p_i^2-m_i^2) a_i =1 \quad, \label{cons-2}\\
\kappa^\mu=0 \quad. \label{cons-3}
\end{align}
Most importantly, we should find a prerequisite for the validity of reduction formula assumed in Eq.~\eqref{span0} by
taking a careful look at the condition of Eq.~\eqref{cons-3}:
\begin{align}
\kappa^\mu=p_1^\mu a_1+p_2^\mu a_2+\cdots+p_{n-1}^\mu a_{n-1}=0 \quad.
\end{align}
If a set of vectors, $\{p_i;i=1,\cdots,n-1\}$, are linearly independent, then every coefficients of those vectors,
$a_0,a_1,\cdots,a_{n-1}$, are all zero, which is a trivial solution. In order to have non--trivial solution, the set of those vectors should, therefore, be \emph{linearly dependent}, where the total number of vectors should be greater than the dimension of vector space $D$. Consequently, $n-1\ge D$ must be satisfied. In four--dimensional vector space, $n>5$. (This constraint is not valid for the linearly independent case in Sec~\ref{sec2-2}).

Next, we should consider whether a system of linear equations, Eq.~\eqref{cons-1}, Eq.~\eqref{cons-2}, and Eq.~\eqref{cons-3}, is solvable or not.
Eq.~\eqref{cons-1} is equivalent to one of Eq.~\eqref{cons-3} because Eq.~\eqref{cons-3} doesn't change when we apply Eq.~\eqref{cons-1} to Eq.~\eqref{cons-3}. Therefore, the total number of equations is five while the number of unknowns is $n$.
Consequently $n\ge5$ for the underdetmined system. This means that the reduction formula, Eq.~\eqref{span0} is valid only for
the more than five point scalar integral.

$\kappa^\mu=0$ in Eq.~\eqref{cons-3} can be modified as a scalar equation by dotting $p_\ell$:
\begin{align}\label{cond3}
2p_\ell  \cdot  \kappa =2\sum_{i=0}^{n-1} (p_\ell  \cdot  p_i)  a_i=0 \quad,
\end{align}
where $\ell$ can be one of $\{0,1,\cdots,n-1\}$, and a factor $2$ is introduced for later use.

In an attempt to derive a concise matrix form, let us begin with Eq.~\eqref{cond3}:
\begin{align}
0&=2\sum_{i=1}^{n-1} (p_\ell  \cdot  p_i)  a_i=2p_\ell^2 a_\ell+2\sum_{i=0,i\ne \ell}^{n-1} (p_\ell\.p_i) a_i \nonumber \\
&=p_\ell^2 a_\ell +\left(1+m_\ell^2 a_\ell-\sum_{i=0,i\ne \ell}^{n-1}(p_i^2-m_i^2)a_i\right)+2\sum_{i=0,i\ne \ell}^{n-1} (p_\ell\.p_i) a_i \nonumber \\
&=1+2m_\ell^2 a_\ell -\sum_{i=0,i\ne \ell}^{n-1}\left(p_i^2-m_i^2+p_\ell^2-m_\ell^2 -2 p_\ell\.p_i \right)a_i \nonumber \\
&=1+\sum_{i=0}^{n-1}\left(m_\ell^2+m_i^2-p_{\ell i}^2\right) a_i \quad,\nonumber
\end{align}
where $p_{ij}$ is defined as $p_i-p_j$. To reach the second line from the first, we applied Eq.~\eqref{cons-2}. We also used $\sum a_i=0$.

Therefore, by varying $\ell$ from $0$ to $n-1$:
\begin{align}
-\sum_{j=0}^{n-1} \left(m_i^2+m_j^2-p_{ij}^2\right) a_j=-\sum_{j=0}^{n-1} Y_{ij} \, a_j=1 \quad.
\end{align}
Here, the matrix $Y$ is called Cayley matrix. Now a system of equations becomes:
\begin{align}
-Y\.A= \nonumber -\begin{pmatrix}
2m_0^2                                         & \cdots & m_0^2+m_{n-1}^2-p^2_{0\,n-1} \\
m_1^2+m_0^2-p^2_{1\,0}                & \cdots & m_1^2+m_{n-1}^2-p^2_{1\,n-1} \\
\vdots                                         & \ddots & \vdots                                         \\
m_{n-1}^2+m_0^2-p^2_{n\!-\!1\,\,0}  & \cdots & 2m_{n-1}^2
\end{pmatrix}
\begin{pmatrix}
a_0 \\
a_1 \\
\vdots \\
a_{n-1}
\end{pmatrix}
=\begin{pmatrix}
1 \\
1 \\
\vdots \\
1
\end{pmatrix} \quad.
\end{align}
If a determinant, $|Y|$, does not vanish, the above matrix equation can be completely solved for the $a_i$ coefficients.
Consequently, the $n$-point scalar integral satisfies the following reduction formula:
\begin{align}\label{master1}
T^n_0 = \sum_{i=0}^{n-1} a_i \, T^{n-1}_0(i), \quad n > 5 \quad,
\end{align}
with
\begin{align}\label{master2}
a_i=-\sum_{j=0}^{n-1} \left(Y^{-1}\right)_{ij},\,\,\,\,(i\,=0,\cdots,n-1) \quad.
\end{align}
This result is exactly identical to that of Ref.~\cite{Denner:2005nn} for $n$-point scalar integrals with $n\ge 6$.

For discussion, let us consider $6$--point scalar integral.
The reduction formula is based on solving a system of equations including $\kappa^\mu=0$.
\begin{align*}
\kappa^\mu=a_1 p_1^\mu+a_2 p_2^\mu+a_3 p_3^\mu+a_4 p_4^\mu+a_5 p_5^\mu=0 \,\,.
\end{align*}
Physically and mathematically, it is always true that any five vectors are linearly dependent
in a four--dimensional vector space if all of five momenta does not vanish.
Difficulties would appear when, for instance, all components of $p_1^\mu$ are so small to be negligible,
which may correspond to soft gluon radiation from a hexagon diagram; in this specific case, $p_2^\mu,\cdots,p_5^\mu$
could be linearly independent so that reduction formula, Eq.~\eqref{master1} is not safe. The remedy of which will be provided in Sec.~\ref{sec2-2}. (The resultant expression is identical).

\subsection{\label{sec2-2}For the case of linear independence on external momenta}

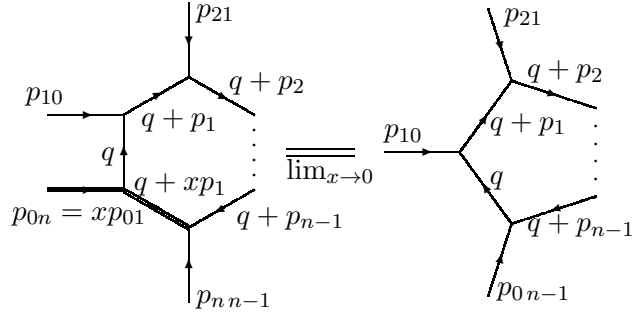
\begin{figure}
\begin{center}
\begin{picture}(84,50)
\qbezier(25,10)(25,10)(33.66,15)
\qbezier(25,10.2)(25,10.2)(16.34,15.2)
\qbezier(25,9.7)(25,9.7)(16.34,14.7)
\put(16.34,15){\line(0,1){10}}
\put(16.34,20){\vector(0,1){1}}
%\put(38.66,15){\line(0,1){10}}
\qbezier(25,30)(25,30)(16.34,25)
\qbezier(25,30)(25,30)(33.66,25)
\put(25,0){\vector(0,1){6}}
\put(25,6){\line(0,1){4}}
\qbezier(6.34,15.2)(6.34,15.2)(16.34,15.2)
\qbezier(6.34,14.8)(6.34,14.8)(16.34,14.8)
\put(11.34,15){\vector(1,0){1}}
\qbezier(6.34,25)(6.34,25)(16.34,25)
\put(6.34,25){\vector(1,0){6}}
\put(25,40){\vector(0,-1){6}}
\put(25,34){\line(0,-1){4}}
%\qbezier(48.66,30)(48.66,30)(38.66,25)
%\put(48.66,30){\vector(-2,-1){6}}
\put(33.66,25){\circle*{0.5}}
\put(33.66,23){\circle*{0.5}}
\put(33.66,21){\circle*{0.5}}
\put(33.66,19){\circle*{0.5}}
\put(33.66,17){\circle*{0.5}}
\put(33.66,15){\circle*{0.5}}
\put(20.67,12.5){\vector(-2,1){1}}
\put(20.67,27.5){\vector(2,1){1}}
\put(29.33,27.5){\vector(2,-1){1}}
\put(29.33,12.5){\vector(-2,-1){1}}
\put(3.34,27){$p_{10}$}
\put(1.5,11){$p_{0n}=xp_{01}$}
\put(26,0){$p_{n\,n-1}$}
\put(31.33,10.5){$q+p_{n-1}$}
\put(30.33,28.5){$q+p_2$}
\put(26,38){$p_{21}$}
\put(13.34,19){$q$}
\put(18.67,23.5){$q+p_{1}$}
\put(17.67,15){$q+xp_{1}$}
\put(38,20.5){\line(1,0){9}}
\put(38,19.5){\line(1,0){9}}
\put(38,16.5){$\lim_{x \rightarrow 0}$}
\put(51,20){\line(1,0){10}}
\qbezier(61,20)(61,20)(67.91,29.51)
\qbezier(67.91,29.51)(67.91,29.51)(79.09,25.88)
\qbezier(79.09,14.12)(79.09,14.12)(67.91,10.49)
\qbezier(67.91,10.49)(67.91,10.49)(61,20)
\put(79.09,25.88){\circle*{0.5}}
\put(79.09,23.53){\circle*{0.5}}
\put(79.09,21.18){\circle*{0.5}}
\put(79.09,18.83){\circle*{0.5}}
\put(79.09,16.48){\circle*{0.5}}
\put(79.09,14.12){\circle*{0.5}}
\qbezier(64.82,0.98)(64.82,0.98)(67.91,10.49)
\qbezier(64.82,39.02)(64.82,39.02)(67.91,29.51)
\put(56.5,20){\vector(1,0){1}}
\put(66.52,33.79){\vector(1,-4){0.1}}
\put(66.52,6.21){\vector(1,4){0.1}}
\put(64.7,25){\vector(2,3){0.1}}
\put(74,27.6){\vector(3,-1){0.1}}
\put(73.1,12.25){\vector(-3,-1){0.1}}
\put(64.1,15.8){\vector(-2,3){0.1}}
\put(51,22){$p_{10}$}
\put(67,37.5){$p_{21}$}
\put(66,1){$p_{0\,n-1}$}
\put(65,23){$q+p_1$}
\put(70,30){$q+p_2$}
\put(70,9.5){$q+p_{n-1}$}
\put(65,15.8){$q$}
\end{picture}
\end{center}
\caption{Diagrammatic description of the reduction (Eq.~\eqref{masterx}) introduced in Sec.~\ref{sec2-2}. $(n+1)$-point integral reduces
to $n$-point integral. $p_\ell$ is chosen to be $p_1$, and $p_n=p_{n0}$ should be (anti)parallel to $p_\ell$.
In order to avoid unphysical singularities, external legs $p_{10}$, $p_{n0}$ and the first and last loop propagators should share the same mass term.}\label{figBR}
\end{figure}

The linear independence on $p_i$'s prohibits the use of compact formula that is derived at previous section. However, we can introduce $p_n$ as a $n^{\rm th}$ denominator $N_n$ to Eq.~\eqref{ids} in order to break down the independence of $p_1, \cdots, p_{n-1}$:
\begin{align*}
p_n= x p_1 \quad.
\end{align*}
With $N_n$, we can apply the same identity as described in Sec.~\ref{sec2-1}.
\begin{align*}
\frac{1}{N_0\cdots N_{n-1}N_n}=\frac{\sum_{i=0}^{n-1} a_i \, N_i }{N_0\cdots N_{n-1}N_n}\quad, \qquad \forall q^\mu \quad,
\end{align*}
where $a_i=-\sum_{j=0}^n \left(Y^{\,-1}\right)_{ij}$. Multiplying $q^2-m_0^2$ in both sides gives:
\begin{align}\label{ids1}
\frac{q^2-m_0^2}{N_0\cdots N_{n-1}N_n}=\frac{\sum_{i=0}^{n-1} (q^2-m_0^2)a_i N_i }{N_0\cdots N_{n-1}N_n}\quad, \qquad \forall q^\mu \quad.
\end{align}

While $x=0$ gives a vanishing Cayley determinant for the $(n+1)$-point loop, the limiting case of $x\rightarrow 0$ reduces this $(n+1)$-point
into $n$-point integral as shown in Fig.~\ref{figBR}. For small $x$, the unphysical external leg of $p_{0n}(=xp_{01})$ can be absorbed by the external leg of $p_{10}$ when mass term is shared. Besides, a loop propagator of $q+xp_1$ can also be absorbed by the external leg of $q$ with the shared mass term. The fact that the fermion current is always involved in a loop guarantees the existence of a common mass term.
The identity Eq.~\eqref{ids1} supposes that a fermion current consists of $p_{nn-1}$, $q$ and $p_{10}$.
The mass of the particle on an external leg of $p_{0n}$ is assumed to be $m_0$, which is also the mass of $n^{\rm th}$ propagator $N_n$ as illustrated in Fig.~\ref{figBR}.
Employing the limit $x\rightarrow 0$ takes it back to the result of the linearly dependent case in Sec.~\ref{sec2-1}:
\begin{align}
T^{n}&=\lim_{x\rightarrow 0} \int_{D} \frac{q^2-m_0^2}{N_0\cdots N_{n-1}N_n}=\lim_{x\rightarrow 0} \int_{D}
\frac{\sum_{i=0}^{n-1} (q^2-m_0^2)a_i N_i }{N_0\cdots N_{n-1}N_n}\nonumber \\
&=\lim_{x\rightarrow 0} \int_{D} \frac{\sum_{i=0}^{n-1} a_i N_i }{N_0\cdots N_{n-1}}=\sum_{i=0}^{n-1} a_i T^{n-1}(i) \quad.
\end{align}
Hence, the previous reduction formula described in Sec.~\ref{sec2-1} is now extended to the region where external momenta are linearly independent.

\begin{figure}
\begin{center}
\includegraphics[width=0.5\textwidth]{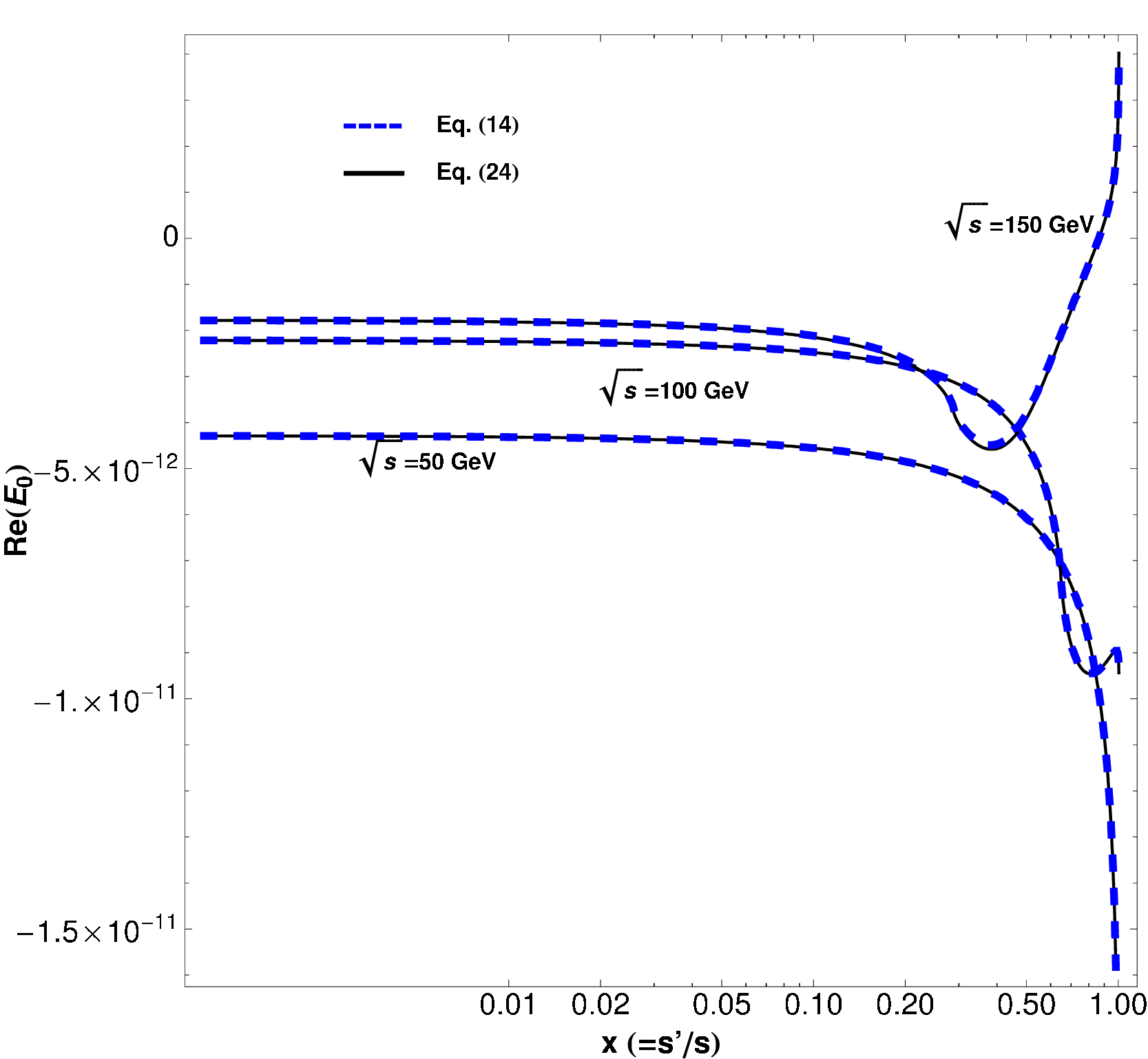}
\caption{\label{figE}Comparison between Eq.~\eqref{master1} and Eq.~\eqref{masterx} for five-point scalar integral ($E_0$)
at different energy scales ($\sqrt{s}$). Those two reductions generate identical numerical results
which show the reduction formula described in Sec.~\ref{sec2-1} can be extended to the region where momenta are linearly independent. The Feynman diagram for this computation is illustrated at Fig.~\ref{fdEF}.}
\end{center}
\end{figure}

In order to verify the reduction procedure illustrated in Fig.~\ref{figBR} \emph{numerically}, we should expand to second-order in $x$.
For this goal, let us start from vector integrals:
\begin{align}\label{indv}
T^n + 2b_\ell \sum_{i=1}^{n-1} p_\ell \. p_i T^{n}_i =\sum_{i=0}^{n-1} a_i T^{n-1}(i) \quad.
\end{align}
This reduction can be easily derived by considering the equation as follows:
\begin{align}\label{indidv}
\frac{2b_\ell q\. p_\ell +1}{N_0\cdots N_{n-1}}=\frac{\sum_{i=0}^{n-1} a_i N_i}{N_0\cdots N_{n-1}} \quad.
\end{align}
Coefficients $a_i$'s and $b_\ell$ must be solutions of a system of equations shown below:
\begin{align*}
\sum_{i=0}^{n-1} a_i =0 \,\,, \qquad \sum_{i=0}^{n-1} (p_i^2-m_i^2) a_i =1\,\, , \qquad
a_1 p_1^\mu +\cdots a_{n-1} p_{n-1}^\mu=b_\ell p_\ell \label{indeq}\,\, .
\end{align*}
The fact that the $p_i$'s are linearly independent implies that all $a_i$'s vanish except $a_\ell$ and $b_\ell$:
\begin{align*}
a_0+b_\ell =0 \quad, \qquad -m_0^2 a_0+(p_\ell^2-m_\ell^2) a_\ell=1  \quad,
\end{align*}
which gives:
\begin{align}
b_\ell=a_\ell=-a_0=\frac{1}{f_\ell} \quad.
\end{align}
With these coefficients, and Eq.~\eqref{indidv}, we can derive a reduction formula for the vector integral by varying $\ell=1,\cdots,n-1$:
\begin{align}\label{indvec}
T^n_i=\sum_{j=1}^{n-1}\left(Z^{-1}\right)_{ij} \left[T^{n-1}(j)-T^{n-1}(0)-f_j T^n\right] \quad,
\end{align}
where the Gram matrix, $Z$, is defined as $Z_{ij}=2 p_i \. p_j$, and $f_i=p_i^2-m_i^2+m_0^2$. This reduction formula is exactly the same form of the conventional Parssarino-Veltman reduction.

For later use, the identity, Eq.~\eqref{indidv}, leads to:
\begin{align}\label{veckey}
\frac{2 q\.p_\ell}{N_0\cdots N_{n-1}}=\frac{N_\ell}{N_0\cdots N_{n-1}}-\frac{N_0}{N_0\cdots N_{n-1}}-\frac{f_\ell}{N_0\cdots N_{n-1}} \,\,.
\end{align}
Now, we are ready to return to the scalar integral. By means of multiplying Eq.~\eqref{ids} by $(q^2-m_0^2)$ after the shift of $n\rightarrow n+1$, the identity for the $T^{n+1}$ scalar integral is:
\begin{align}\label{indids0}
\frac{q^2-m_0^2}{N_0 \cdots N_n}=\sum_{i=0}^{n} \frac{(q^2-m_0^2) a_i N_i}{N_0\cdots N_n} \quad.
\end{align}
This is the same form of the identity of Eq.~\eqref{ids1}. Multiplying both sides with $N_n/(q^2-m_0^2)$ gives:
\begin{align*}
\frac{1}{N_0 \cdots N_{n-1}}&=\sum_{i=1}^{n-1} \frac{a_i N_i}{N_0\cdots N_{n-1}}+\frac{a_0 N_0}{N_0\cdots N_{n-1}}
+\frac{a_n (q^2+x^2p_\ell^2+2 x q\. p_\ell-m_0^2)}{N_0\cdots N_{n-1}}\\
&=\sum_{i=1}^{n-1} \frac{a_i N_i}{N_0\cdots N_{n-1}}+\frac{(a_0+a_n) N_0}{N_0\cdots N_{n-1}} +\frac{a_n (x^2p_\ell^2)}{N_0\cdots N_{n-1}}
+\frac{x a_n (2 q\. p_\ell)}{N_0\cdots N_{n-1}} \quad.
\end{align*}
Taking integration with the limit $x\rightarrow 0$, and the help of Eq.~\eqref{veckey}, leads to the reduction formula:
\begin{align}
T^n=\sum_{i=0}^{n-1} b_i T^{n-1}(i)\quad, \label{masterx}
\end{align}
where the $b_i$'s are:
\begin{align*}
b_i=\lim_{x\rightarrow0}\frac{a_i + \delta_{i\ell} x a_n + \delta_{i0}(1-x) a_n}{1-a_n (x^2p_\ell^2-xf_\ell)} \quad.
\end{align*}
Numerical computation of the above reduction formula is identical to the general reduction formula of Eq.~\eqref{master1} as shown in Fig.~\ref{figE}.
\begin{figure}
\begin{center}
\includegraphics[width=0.35\textwidth]{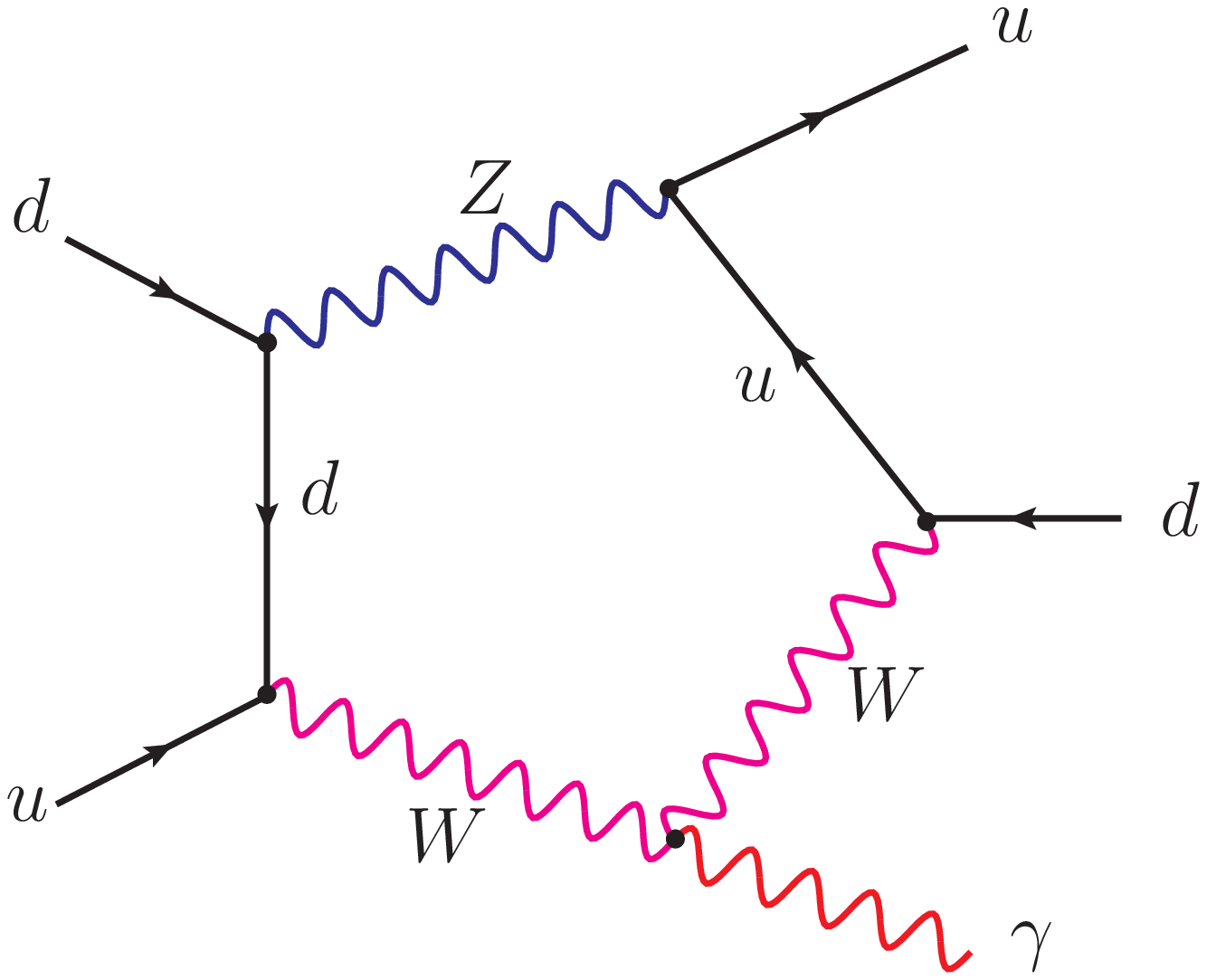}\hskip 1.5cm%
\includegraphics[width=0.35\textwidth]{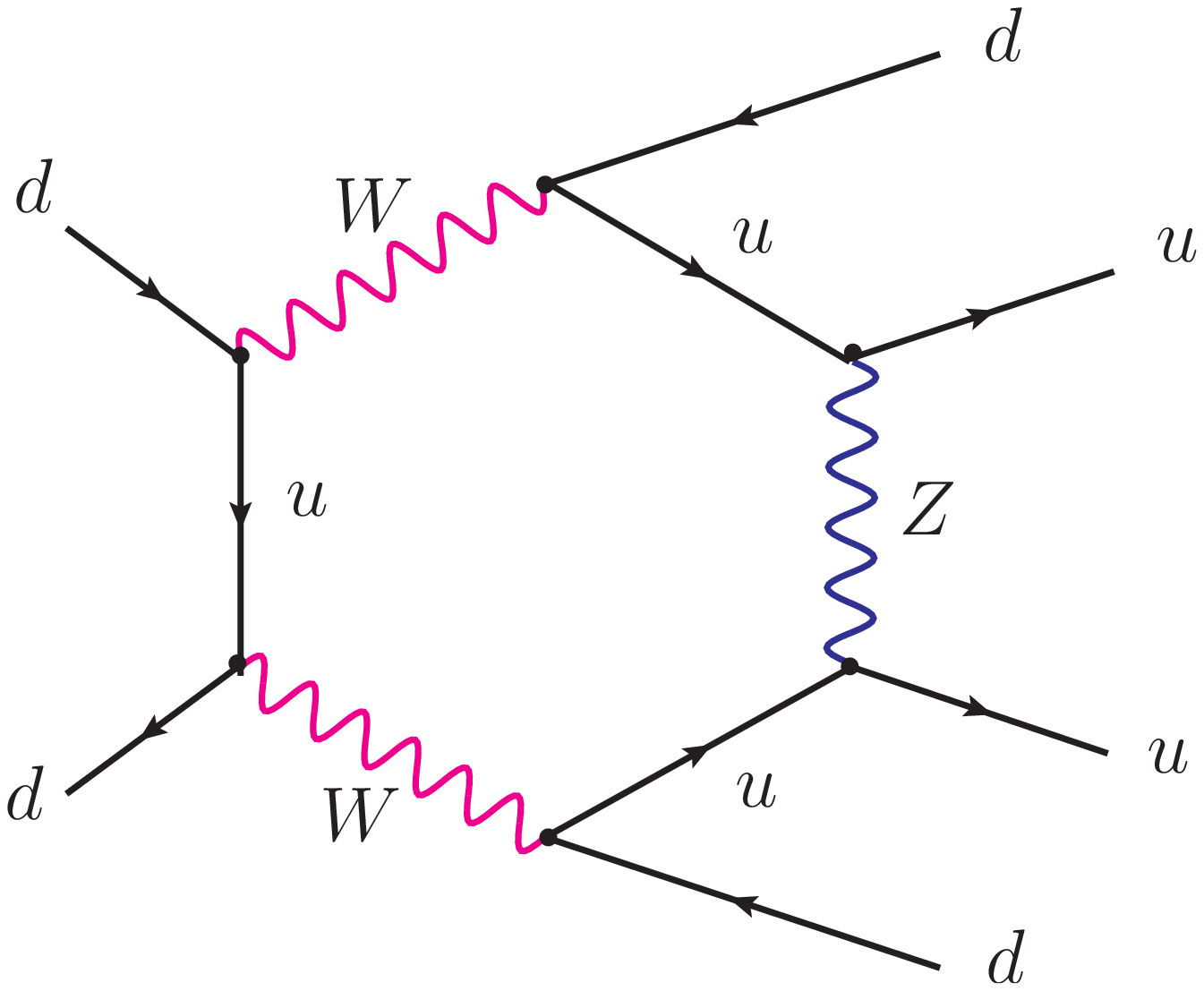}%
\caption{\label{fdEF}One-loop pentagon and hexagon diagrams for the computation of five ($E_0$) and six ($F_0$)-point scalar integrals, which are drawn by {\it JaxoDraw}~\cite{Binosi:2008ig}}
\end{center}
\end{figure}

\subsection{\label{sec2-3}For the one-loop vector and tensor $n$-point integrals}

Reduction for the vector and tensor integrals can be determined without difficulty. Although we multiply Eq.~\eqref{ids} and Eq.~\eqref{ids1} with $q^{\mu_1} \cdots q^{\mu_r}$ on both sides, the identity is still valid for all $q^\mu$. Hence, the following formula can be obtained:
\begin{align}\label{red}
T^n_{\mu_1\cdots \mu_r}&=
\frac{(2 \pi \mu)^{4-D}}{i\pi^2} \int d^D q \frac{q_{\mu_1} \cdots q_{\mu_r}}{N_0 N_1 \cdots N_{n-1}} \nonumber \\
&=\sum_{i=0}^{n-1} a_i T^{n-1}_{\mu_1\cdots \mu_r}(i) \quad.
\end{align}

Tensor integrals can be decomposed to scalar coefficient functions:
\begin{align}\label{decomp}
\left(T^n\right)^{\mu_1 \cdots \mu_r}=\sum_{k=0}^{[r/2]} \sum_{i_1,i_2,\cdots,i_r=1}^{n-1} \left\{
\begin{matrix} p_{i_1}^{\mu_1}\cdots p_{i_r}^{\mu_r}\\2k\end{matrix}\right\} T^n_{\underbrace{ \scriptscriptstyle{00\cdots00}}_{2k} i_1\cdots i_{r-2k}} \quad,
\end{align}
where we used combitoric notation $\left\{ \begin{smallmatrix} p_{i_1}^{\mu_1}\cdots p_{i_r}^{\mu_r}\\2k\end{smallmatrix}\right\}$,
which denotes that; First, pick out $2k$ out of $\{p_{i_1}^{\mu_1}\cdots p_{i_r}^{\mu_r}\}$. Second, switch picked--out $p^\mu p^\nu \cdots$
to $g^{\mu \nu}g^{..}\cdots$ as a sum of all possible combinations of Lorenz indices followed by leftovers of $p$'s. Finally, the sub--indices of momentum $i$ must be kept same order starting from $i_1$ to $i_{r-2k}$ which is just dummy indices for summation. $[r/2]$ is the maximal integer lass than or equal to $r/2$. Note that we apply possible combinations only for Lorenz indices, not for the momentum subscripts. 
As examples, two--point functions are decomposed:
\begin{align*}
\left\{ \begin{matrix} p_{i_1}^{\mu}p_{i_2}^{\nu}p_{i_3}^{\rho}\\2\end{matrix}\right\}=g^{\mu \nu} p^\rho_{i_1}+g^{\mu \rho} p^\nu_{i_1}+g^{\nu \rho} p^\mu_{i_1}\quad,
\end{align*}
which gives:
\begin{align}\label{Bs}
B_\mu&=p_{1 \mu} B_1, \nonumber \\
B_{\mu \nu}&=p_{1\mu}p_{1\nu} B_{11}+g_{\mu \nu}B_{00}, \nonumber \\
B_{\mu \nu \rho}&=p_{1\mu}p_{1\nu}p_{1\rho}B_{111}+\left(g_{\mu \nu}p_{1 \rho}+g_{\mu \rho}p_{1 \nu}+g_{\nu \rho}p_{1 \mu}\right)B_{001}, \nonumber \\
B_{\mu \nu \rho \sigma}&=p_{1\mu}p_{1\nu}p_{1\rho}p_{1\sigma}B_{1111}+\left(g_{\mu \rho}g_{\rho \sigma}+g_{\nu \rho}g_{\mu \sigma}+g_{\rho \mu}g_{\nu \sigma}\right)B_{0000} \nonumber \\ &+
\left(g_{\mu \nu}p_{1 \rho}p_{1 \sigma}+g_{\mu \rho}p_{1 \sigma}p_{1 \nu}+g_{\mu \sigma}p_{1 \nu}p_{1 \rho}+g_{\nu \rho}p_{1 \sigma}p_{1 \mu}
+g_{\rho \sigma}p_{1 \nu}p_{1 \mu}+g_{\sigma \nu}p_{1 \rho}p_{1 \mu}\right)B_{0011}.
\end{align}

Using the decomposition of Eq.~\eqref{decomp}, $(T^{n-1})^{\mu_1\cdots\mu_r}(k)$ that is defined in Eq.~\eqref{T(k)} becomes:
\begin{align}\label{decomp1}
\left(T^{n-1}\right)^{\mu_1 \cdots \mu_r}(l \ne 0)&= \quad\quad \sum_{k=0}^{[r/2]}
\sum_{i_1,\cdots,i_r=1}^{n-1} \bar{\delta}_{i_1 l}\cdots\bar{\delta}_{i_rl}\left\{
\begin{matrix} p_{i_1}^{\mu_1}\cdots p_{i_r}^{\mu_r}\\2k\end{matrix}\right\}
T^{n-1}_{\underbrace{ \scriptscriptstyle{00\cdots00}}_{2k} \theta_{i_1l}\cdots \theta_{i_{r-2k}l}}(l) \,\, ,\nonumber \\
\left(T^{n-1}\right)^{\mu_1 \cdots \mu_r}(0)&= \sum_{k=0}^{[r/2]} \sum_{i_1,\cdots,i_r=1}^{n-1} \left\{
\begin{matrix} p_{i_1}^{\mu_1}\cdots p_{i_r}^{\mu_r}\\2k\end{matrix}\right\}  T^{n-1}_{\underbrace{ \scriptscriptstyle{00\cdots00}}_{2k} i_1 \cdots i_{r-2k}}(l) \,\, ,
\end{align}
with,
\begin{align}
\theta_{ij}&=\left\{ \begin{matrix} i & {\rm when} \,\, i<j \\ i-1 & {\rm when} \,\, i>j \end{matrix} \right.\quad, \nonumber \\
\bar{\delta}_{ij}&=1-\delta_{ij}\quad.
\end{align}
We can directly derive the vector and tensor coefficient functions by comparing Eq.~\eqref{decomp} with Eq.~\eqref{decomp1}
since they are connected by the reduction formula of Eq.~\eqref{red}:
\begin{align*}\label{tensorR}
T^n_{\underbrace{ \scriptscriptstyle{00\cdots00}}_{2l} i_1 \cdots i_{r-2l}}=\qquad
a_0T^{n-1}_{\underbrace{ \scriptscriptstyle{00\cdots00}}_{2l} i_1 \cdots i_{r-2l}}(0)+\sum_{k=1}^{n-1}a_k\bar{\delta}_{i_1k}\cdots
\bar{\delta}_{i_1k}T^{n-1}_{\underbrace{ \scriptscriptstyle{00\cdots00}}_{2l} i_1 \cdots i_{r-2l}}(k)\quad.
\end{align*}

\section{\label{sec3}Three--point and four--point functions, C, D. -- for the region of small kinematic variables}
\begin{figure}
\includegraphics[width=0.48\textwidth]{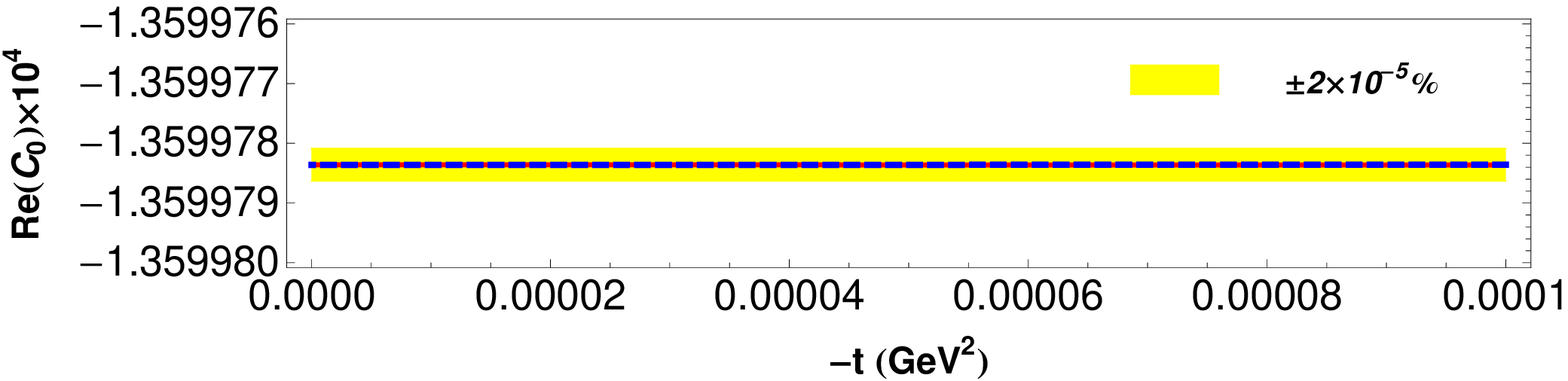}%
\includegraphics[width=0.48\textwidth]{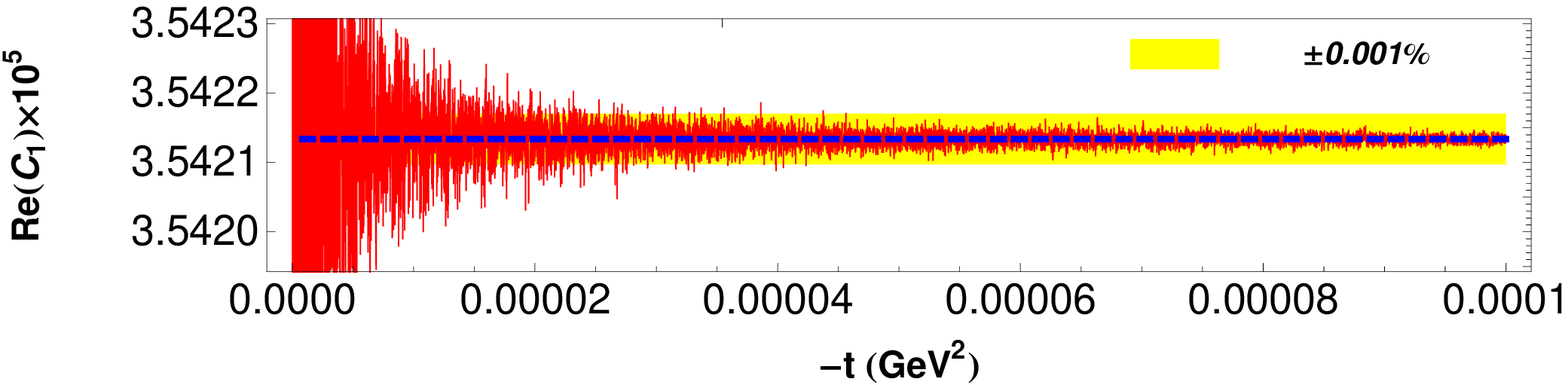}%

\includegraphics[width=0.48\textwidth]{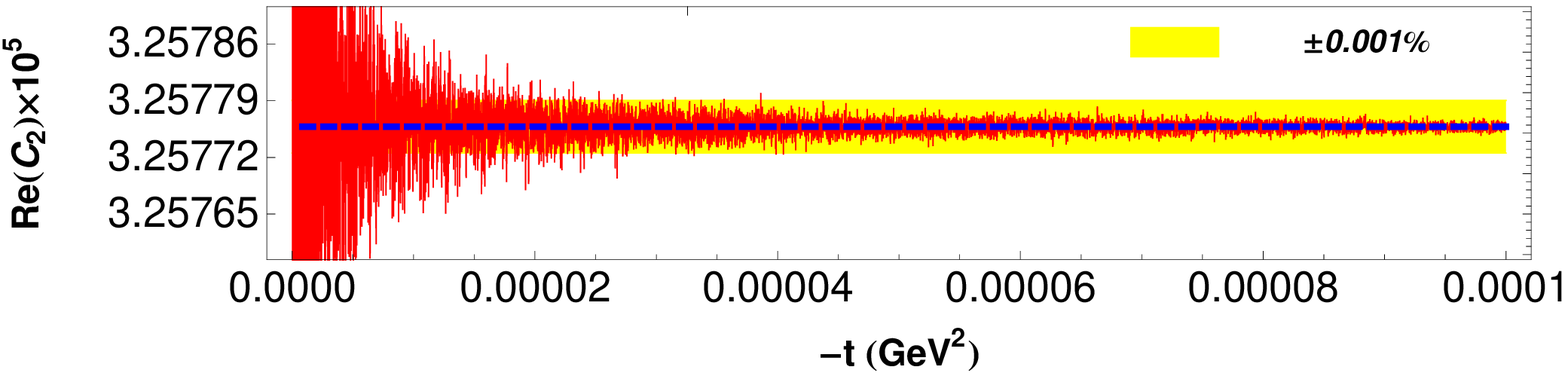}%
\includegraphics[width=0.48\textwidth]{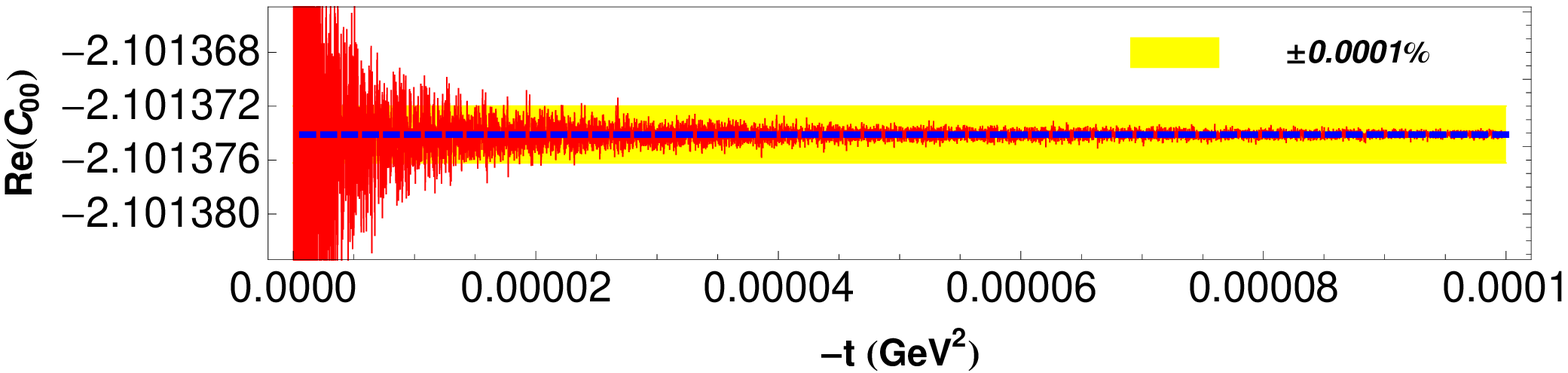}%

\includegraphics[width=0.48\textwidth]{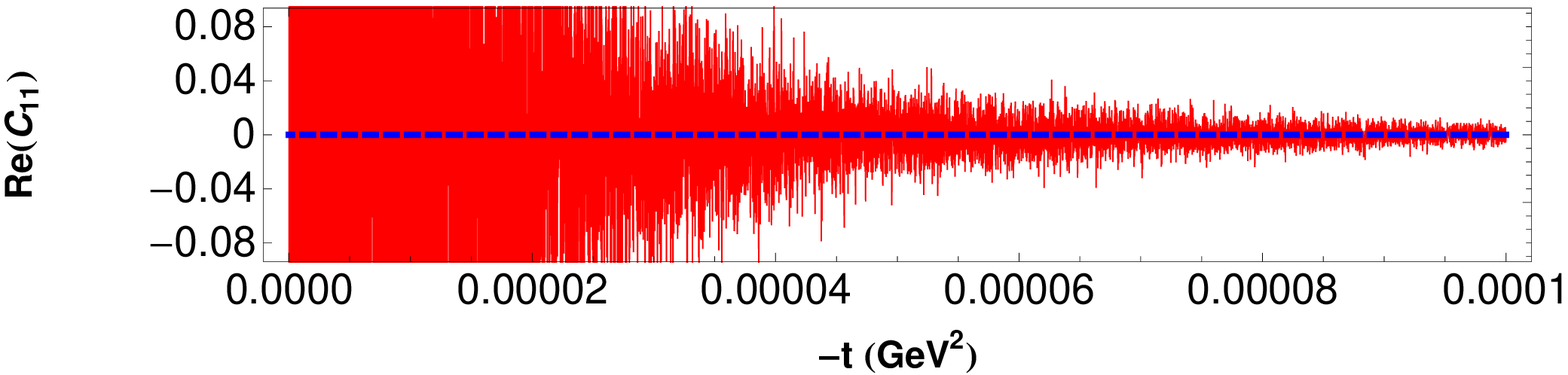}%
\includegraphics[width=0.48\textwidth]{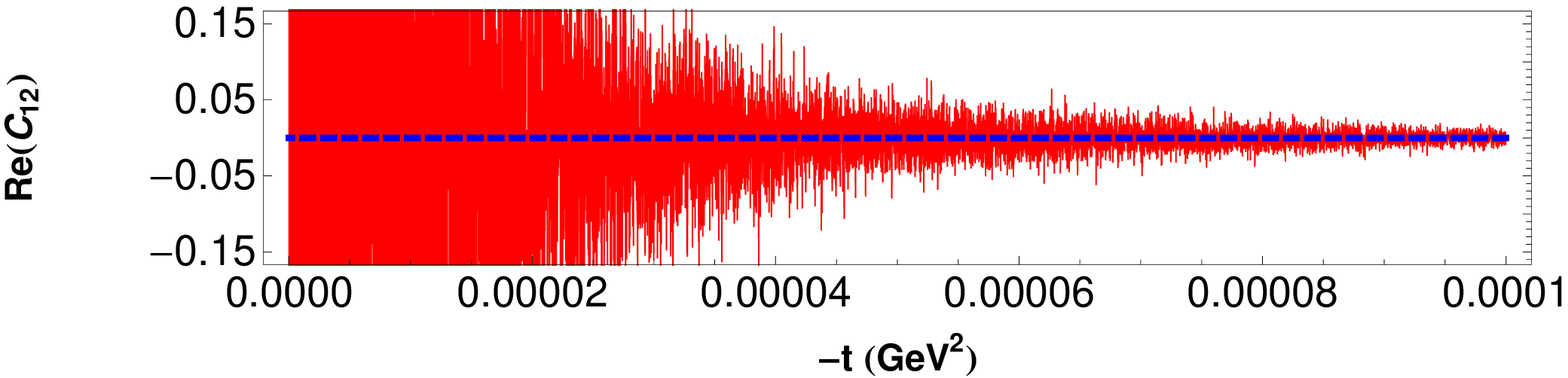}%

\includegraphics[width=0.48\textwidth]{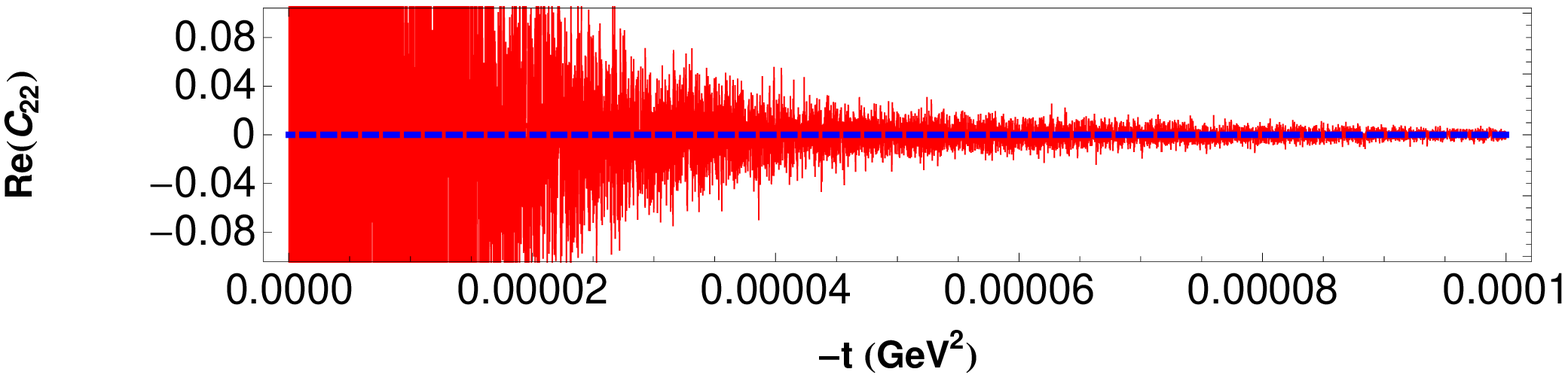}%
\caption{\label{figCsm}Numerical safety of three-point functions : $C_{\cdots}[p_1^2, p_2^2,t,m_0^2, m_1^2, m_2^2]$$=$$C_{\cdots}[0, 0,t,0, M_W^2, M_Z^2]$ with $M_W=80.425\,\,GeV,\,\,\,M_Z=91.1876\,\,GeV$, And $t$ varies from $-0.0001$ to $0$.
The conventional Passarino-Veltman reduction shows numerical instability in small kinematic region (red curve, computed via
 {\it LoopTools}~\cite{Hahn:2000jm}),
while blue curve which generated from the method in Sec.~\ref{sec3} (Eq.~\eqref{Cexpr1},~\eqref{Cexpr2}) is numerically stable. The Feynman diagram for this computation is shown in Fig.~\ref{fdCD}.}
\end{figure}
In Sec.~\ref{sec2}, we studied the algebraic formalism for the reduction of one-loop integrals.
The coefficient functions of four--point or less--point integrals reside in inner--product space where the dimension of vector space is four in general. Hence, the number of momentum involved in the loop is always less than four. This mostly guarantees the dependence of momenta.
Expressions for one-- ($A$), two-- ($B$), three-- ($C$), and four--point  ($D$) functions have already been known in an analytical form~\cite{'tHooft:1978xw,Denner:1991kt}.
However, we need the solutions to fix the problem that appear in the numerical computation,
such as the case of the vanishing Gram determinant. In Ref.~\cite{Denner:2005nn}, the technique of numerical iteration is suggested
for the small Gram determinant.

The method in Sec.~\ref{sec2} provides the analytical approach for the case of small kinematic variables. If $p_i\. p_j = 0$ and $p_i^2= 0$ for all $i$ and $j$, then the identity of Eq.~\eqref{ids} can be greatly simplified.
Furthermore, every $n$-point functions can be reduced to a sum of two-point functions:
\begin{align}
\frac{1}{N_0\cdots N_{n-1}}=\sum_{i=1}^{n-1} \frac{\mathcal{A}_i}{N_0 N_i} \quad.
\end{align}
After some simple algebra, the $\A_i$'s are easily determined:
\begin{align}
\mathcal{A}_i=\frac{1}{\prod_{j=0 \atop j\ne i}^{n-1} (m_i^2-m_j^2)}\quad, \qquad \sum_{i=1}^{n-1} \mathcal{A}_i=0 \quad.
\end{align}

If the $p_i\.p_j$'s are small enough, the loop propagator can be expanded around zero using the small variables,
$\hat{z}=p^2$, $\tilde{z}=2 q\.p$ and $z=\hat{z}+\tilde{z}$:
\begin{align}
\frac{1}{(q+p)^2-m^2}&= \frac{1}{q^2-m^2}-\frac{\hat{z}+\tilde{z}}{(q^2-m^2)^2}+ \frac{(\hat{z}+\tilde{z})^2}{(q^2-m^2)^3}+\cdots \nonumber \\
&=\sum_{k=0}^{\infty} \frac{(-z)^k}{(q^2-k^2)^{k+1}}=\sum_{k=0}^{\infty} \frac{\left(-z/M\right)^k}{M}
=\sum_{k=0}^{\infty} \frac{1}{k!}\left(-z \partial \right)^k(M^{-1}) \,\,,
\end{align}
where the denominator $M=q^2-m^2$ and the differential operator $\partial=\partial/\partial m^2$.
Therefore, with the small values of $\hat{z}_i=p_i^2$, $\tilde{z}_i=2 q\.p_i$, $z_i=\hat{z}_i+\tilde{z}_i$,
the identity of Eq.~\eqref{ids} can be expanded as follows:
\begin{align}\label{smexp}
&\frac{1}{N_0\cdots N_{n-1}}= \left[1-z_i \partial_i+\frac{1}{2!}z_i z_j \partial_i \partial_j
-\frac{1}{3!}z_i z_j z_k \partial_i \partial_j \partial_k +\cdots \right] \frac{1}{M_0\cdots M_{n-1}} \nonumber \\
&=\mathcal{A}_i \frac{1}{M_0M_i}-z_i \partial_i \left(\mathcal{A}_k \frac{1}{M_0M_k} \right)
+\frac{1}{2}z_i z_j \partial_i \partial_j \left(\mathcal{A}_k \frac{1}{M_0M_k} \right)+\cdots\nonumber \\
&= \mathcal{A}_i \frac{1}{M_0M_i}  -z_j \left(\partial_j \mathcal{A}_i\right) \frac{1}{M_0M_i}-z_i \mathcal{A}_i
\left(\partial_i \frac{1}{M_0M_i}\right) +\frac{1}{2}z_j z_k   \left(\partial_j\partial_k\mathcal{A}_i\right) \frac{1}{M_0M_i}\nonumber \\
&\qquad
+z_i z_j   \left(\partial_j\mathcal{A}_i\right) \left(\partial_i\frac{1}{M_0M_i} \right)+\frac{1}{2}z_i^2  \mathcal{A}_i
\left(\partial_i^2 \frac{1}{M_0M_i} \right)+\cdots \nonumber \\
&=\mathcal{A}_i \frac{1}{N_0N_i}-\left(\partial_j \mathcal{A}_i \right)(\hat{z}_j + \tilde{z}_j)\frac{1}{N_0N_i}
+\frac{1}{2!} \left(\hat{z}_j\hat{z}_k+2\hat{z}_j\tilde{z}_k+\tilde{z}_j\tilde{z}_k\right)\left(\partial_j\partial_k
\mathcal{A}_i \right)\frac{1}{N_0N_i} \nonumber \\
&\qquad -\frac{1}{3!} \left(\hat{z}_j\hat{z}_k\hat{z}_\ell+3\hat{z}_j\hat{z}_k\tilde{z}_\ell
+3\hat{z}_j\tilde{z}_k\tilde{z}_\ell+\tilde{z}_j\tilde{z}_k\tilde{z}_\ell\right) \left(\partial_j\partial_k\partial_\ell \mathcal{A}_i \right)
\frac{1}{N_0N_i} +\cdots
\nonumber \\&
=2^0\left(\frac{1}{0!}\mathcal{A}_i -\frac{1}{1!} \hat{z}_{j}\left(\partial_j \mathcal{A}_i \right)
+\frac{1}{2!} \hat{z}_{j}\hat{z}_{k}\left(\partial_j\partial_k \mathcal{A}_i \right)
-\frac{1}{3!}\hat{z}_{j} \hat{z}_{k}\hat{z}_{\ell}\left(\partial_j\partial_k\partial_\ell \mathcal{A}_i \right)
+\cdots \right)\frac{1}{N_0N_i} \nonumber \\
&\qquad +2^1\left(- \frac{1}{1!} \left(\partial_j \mathcal{A}_i \right)+\frac{2}{2!}\hat{z}_{k}\left(\partial_j\partial_k \mathcal{A}_i \right)
-\frac{3}{3!}\hat{z}_{k}\hat{z}_{\ell}\left(\partial_j\partial_k\partial_\ell \mathcal{A}_i \right)
+\cdots \right)p_j^\mu\frac{ q_\mu}{N_0N_i} \nonumber \\
&\qquad +2^2\left( \frac{1}{2!}\left(\partial_j\partial_k \mathcal{A}_i \right)-\frac{3}{3!}\hat{z}_{\ell}
\left(\partial_j\partial_k\partial_\ell \mathcal{A}_i
\right)p_j^\mu +\cdots \right)p_j^\mu p_k^\nu  \frac{q_\mu q_\nu}{N_0N_i} +\cdots \nonumber \\
&
={}^n\!\!\mathcal{F}_{i,(0)} \frac{1}{N_0N_i} +{}^n\!\!\mathcal{F}_{i,(1)}^{\mu} \frac{ q_\mu}{N_0N_i}
+{}^n\!\!\mathcal{F}_{i,(2)}^{\mu \nu} \frac{q_\mu q_\nu}{N_0N_i} +\cdots \,\, ,
\end{align}
where repeated sub-indices imply the summation from $1$ to $n-1$, and coefficient functions $\mathcal{F}$ are defined implicitly.

By integrating Eq.~\eqref{smexp}, the $n$-point scalar integral can be expressed in terms of two--point functions:
\begin{align}\label{mst}
&T^n_0={}^n\!\!\mathcal{F}_{i,(0)} B_0^i +{}^n\!\!\mathcal{F}_{i,(1)}^{\mu} B_\mu^i
+{}^n\!\!\mathcal{F}_{i,(2)}^{\mu \nu} B_{\mu \nu}^i +\cdots \quad,
\end{align}
where,
\begin{align}\label{bis}
B_{\cdots}^i=B_{\cdots}(p_i^2,m_0^2,m_i^2) \quad.
\end{align}
Vector and second rank tensor coefficient functions, $T^n_i$, can be obtained from Eq.~\eqref{smexp} by multiplying $q_\mu$ or $q_\mu q_\nu$:
\begin{align}
&T^n_\mu
={}^n\!\!\mathcal{F}_{i,(0)} B_\mu^i +{}^n\!\!\mathcal{F}_{i,(1)}^{\nu} B_{\mu \nu}^i
+{}^n\!\!\mathcal{F}_{i,(2)}^{\nu \rho} B_{\mu \nu \rho}^i +\cdots  \nonumber \\
&T^n_{\mu \nu}
={}^n\!\!\mathcal{F}_{i,(0)} B_{\mu \nu}^i +{}^n\!\!\mathcal{F}_{i,(1)}^{\rho} B_{\mu \nu \rho}^i
+{}^n\!\!\mathcal{F}_{i,(2)}^{\rho \sigma} B_{\mu \nu \rho \sigma}^i +\cdots \quad.
\end{align}

When two masses are identical (say, $m_a=m_b$), it is necessary to take an additional derivative with respect to $m_a^2$ in Eq.~\eqref{smexp}:
\begin{align}
&\frac{1}{M_0\cdots M_a^2 \cdots  M_{n-2}} = -\partial_a \left(\frac{1}{M_0\cdots M_a \cdots  M_{n-2}}\right)\quad.
\end{align}
Similarly, if $n$ masses are the same, we should take $n$-th order of derivative for the mass.

In Appendix~\ref{ap1}, we present the explicit expressions of the $B$--functions and their derivatives.

In Eq.~\eqref{mst}, $n$--point functions are expressed by two--point ($B$) functions. Alternatively, we can expand Eq.~\eqref{smexp} in terms of three--point ($C$) functions when two masses are identical. As an example, four--point integrals with $m_2=m_3$ for the case of a small Gram determinants yields:
\begin{align*}
\frac{1}{N_0\cdots N_{3}}&=\frac{1}{N_3} \left[ \frac{1}{N_0\cdots N_{2}} \right] \\
&= \frac{1}{N_3}\left[ {}^3\!\!\mathcal{F}_{i,(0)} \frac{1}{N_0N_i} +{}^3\!\!\mathcal{F}_{i,(1)}^{\mu} \frac{ q_\mu}{N_0N_i}
+{}^3\!\!\mathcal{F}_{i,(2)}^{\mu \nu} \frac{q_\mu q_\nu}{N_0N_i} +\cdots \right] \\
&={}^3\!\!\mathcal{F}_{i,(0)} \frac{1}{N_0N_iN_3} +{}^3\!\!\mathcal{F}_{i,(1)}^{\mu} \frac{ q_\mu}{N_0N_iN_3}
+{}^3\!\!\mathcal{F}_{i,(2)}^{\mu \nu} \frac{q_\mu q_\nu}{N_0N_iN_3} +\cdots \quad.
\end{align*}
Therefore,
\begin{align*}
D_0=T^4={}^3\!\!\mathcal{F}_{i,(0)} C_0^i +{}^3\!\!\mathcal{F}_{i,(1)}^{\mu} C_\mu^i
+{}^3\!\!\mathcal{F}_{i,(2)}^{\mu \nu} C_{\mu \nu}^i +\cdots \quad,
\end{align*}
where, 
\begin{align*}
C_{\cdots}^i=C_{\cdots}[p_i^2,p_3^2,p_{i3}^2,m_0^2,m_i^2,m_3^2]\quad.
\end{align*}

\begin{figure}
\begin{center}
\includegraphics[width=0.35\textwidth]{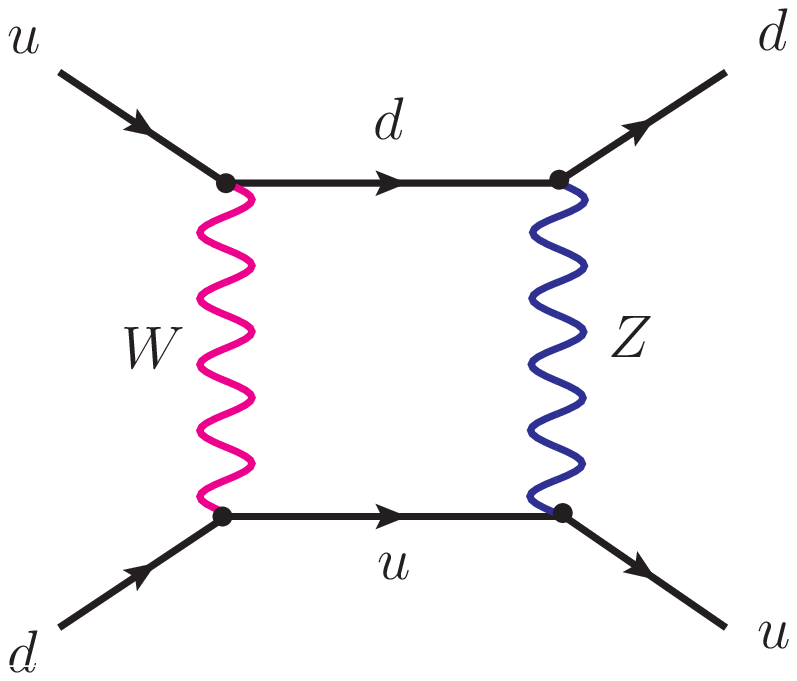}\hskip1.5cm%
\includegraphics[width=0.35\textwidth]{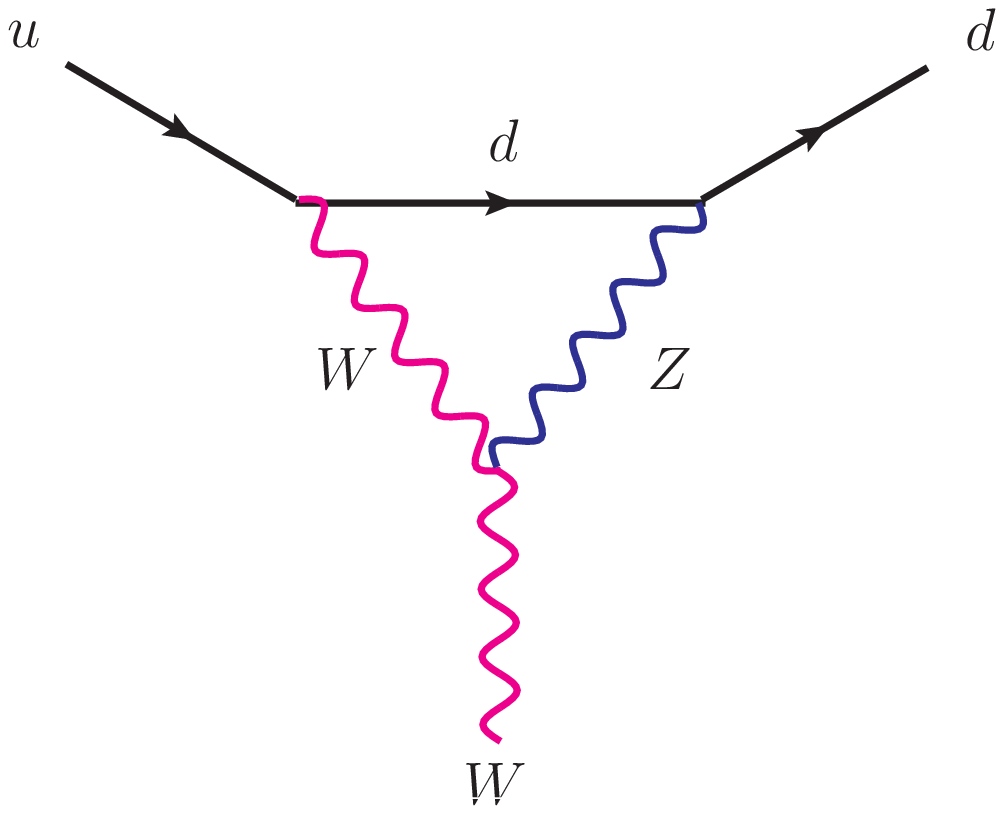}%
\caption{\label{fdCD}One-loop box and triangle diagrams for the computation of four ($D$) and three ($C$)-point scalar integral, which are drawn by {\it JaxoDraw}~\cite{Binosi:2008ig} }
\end{center}
\end{figure}

For the three--point function in the region of small Gram determinants, the $\A_i$ coefficients are:
\begin{align}
\A_1=\frac{1}{m_1^2-m_2^2}\,\, , \qquad \A_2=-\A_1=-\frac{1}{m_1^2-m_2^2} \,\, .
\end{align}
We also introduce the following variables to facilitate compact expressions:
\begin{align*}
\chi_{ij}=\frac{p_j^2-p_i^2}{m_i^2-m_j^2}\,\, , \qquad  \mathcal{P}_{ij}^\mu=\frac{p_j^\mu-p_i^\mu}{m_i^2-m_j^2}\,\, .
\end{align*}

The zeroth order ${}^3\mathcal{F}_{i,(0)}$ can be obtained from the summation of a geometric series with the convergence condition
 $|\chi_{12}|=|\chi_{21}|<1$:
\begin{align}
{}^3\F_{1,(0)}&= \A_1 \left[1-\chi_{12}+\chi_{12}^2 -\chi_{12}^3+\cdots \right]=\frac{\A_1}{1+\chi_{12}}=\omega \quad, \nonumber \\
{}^3\F_{2,(0)}&= \A_2 \left[1-\chi_{21}+\chi_{21}^2 -\chi_{21}^3+\cdots \right]=\frac{\A_2}{1+\chi_{21}}=-\omega \quad,
\end{align}
where $\chi_{ij}=\chi_{ji}$ and $\A_1=-\A_2$ are used.
The first order ${}^3\mathcal{F}_{i,(1)}$ is determined by using the derivative of the geometric series,
\begin{align}
{}^3\mathcal{F}_{1,(1)}&=2\A_1 \left(\partial_{\chi_{12}} \frac{1}{1+\chi_{12}}\right) \P_{12}^\mu =2\omega^2 (p_1^\mu-p_2^\mu) \quad, \nonumber \\
{}^3\mathcal{F}_{2,(1)}&= -2\omega^2 (p_1^\mu-p_2^\mu) \quad.
\end{align}
The $n$-th order derivative of the geometric series provides the $n$-th order ${}^3\F_{i,(n)}$:
\begin{align}
{}^3\F_{1,(n)}&= 2^n    \omega^{n+1} (p_1-p_2)^{\mu_1}\cdots (p_1-p_2)^{\mu_n} \quad,\\
{}^3\F_{2,(n)}&= -2^n  \omega^{n+1} (p_1-p_2)^{\mu_1}\cdots (p_1-p_2)^{\mu_n} \quad.
\end{align}
Up to third order, $C_0$ can be compactly expressed for the case of small kinematic variables as:
\begin{align}\label{Cexpr1}
&C_0=\omega (B^1_0-B^2_0) +2 \omega^2 \left[ (p_1^2-p_1\.p_2) B_1^1+(p_2^2-p_1\.p_2) B_1^2\right] \nonumber \\
&\,\,+4\omega^3 \left[ (p_1-p_2)^2 (B_{00}^1-B_{00}^2)+(p_1^2-p_1\.p_2)^2 B_{11}^1-(p_2^2-p_1\.p_2)^2 B_{11}^2
\right] \nonumber \\
&\,\,+8\omega^4 \big[ 3(p_1-p_2)^2(p_1^2-p_1\.p_2) B_{001}^1+(p_1^2-p_1\.p_2)^3 B_{111}^1 \nonumber \\
&\,\,+3(p_1-p_2)^2(p_2^2-p_1\.p_2) B_{001}^2+(p_2^2-p_1\.p_2)^3 B_{111}^2 \big]
+\mathcal{O}\left((p_1^2)^4,(p_1^2)^4,(p_1\.p_2)^4\right) \quad,
\end{align}
where $B^i_{\cdots}$ is defined in Eq.~\eqref{bis}. 
Similarly, vector and $2^{\rm nd}$-rank-tensor coefficients functions yield:
\begin{align}
&C_1= \omega B^1_1 +2 \omega^2 \left[  B_{00}^1-B_{00}^2+(p_1^2-p_1\.p_2)B_{11}^1 \right] \nonumber \\
&\,\,+4\omega^3 \left[ (3p_1^2-4 p_1\.p_2 +p_2^2)  B_{001}^1 +(p_1^2-p_1\.p_2)^2 B_{111}^1
+2(p_2^2-p_1\.p_2)B_{001}^2
\right] \nonumber \\
&\,\,+8\omega^4 \big[ 3(p_1\!-\!p_2)^2 (B_{0000}^1\!-\!B_{0000}^2)\!+\!6(p_1^2\!-\!p_1\.p_2)^2 B_{0011}^1
\!-\!6(p_2^2\!-\!p_1\.p_2)^2 B_{0011}^2\!+\!(p_1^2\!-\!p_1\.p_2)^3 B_{1111}^1 \big]
\nonumber \\
&\,\,+\mathcal{O}\left((p_1^2)^4,(p_1^2)^4,(p_1\.p_2)^4\right)  \quad,\nonumber \\
\\
&C_2= -\omega B^2_1 +2 \omega^2 \left[ B_{00}^2-B_{00}^1+(p_2^2-p_1\.p_2)B_{11}^2 \right] \nonumber \\
&\,\,-4\omega^3 \left[ (3p_2^2 -4 p_1\.p_2 +p_1^2) B_{001}^2 +(p_2^2-p_1\.p_2)^2 B_{111}^2+2(p_1^2-p_1\.p_2) B_{001}^1
\right] \nonumber \\
&\,\,\!+\!8\omega^4 \big[ 3(p_1\!-\!p_2)^2 (B_{0000}^2\!-\!B_{0000}^1)\!+\!6(p_2^2\!-\!p_1\.p_2)^2 B_{0011}^2\!-\!6(p_1^2\!-\!p_1\.p_2)^2 B_{0011}^1\!
+\!(p_2^2\!-p\!_1\.p_2)^3 B_{1111}^2 \big]
\nonumber \\
&\,\,+\mathcal{O}\left((p_1^2)^4,(p_1^2)^4,(p_1\.p_2)^4\right) \quad, \nonumber
\end{align}

\begin{align}
&C_{00}= \omega ( B_{00}^1 - B_{00}^2 )+2 \omega^2 \left[ (p_1^2-p_1\.p_2)B_{001}^1+ (p_2^2-p_1\.p_2)B_{001}^1\right] \nonumber \\
&\,\,+4\omega^3 \big[ (p_1-p_2)^2 B_{0000}^1+ (p_1^2-p_1\.p_2)^2  B_{0011}^1 - (p_1-p_2)^2B_{0000}^2
-(p_2^2-p_1\.p_2)^2 B_{0011}^2 \big] \nonumber \\
&\,\,+8\omega^4 \big[ 3  (p_1-p_2)^2 (p_1^2-p_1\.p_2) B_{00001}^1 + (p_1^2-p_1\.p_2)^3  B_{00111}^1 \nonumber
\\ & \,\,
+3 (p_1-p_2)^2 (p_2^2-p_1\.p_2) B_{00001}^2 +(p_2^2-p_1\.p_2)^3 B_{00111}^2 \big]
+\mathcal{O}\left((p_1^2)^4,(p_1^2)^4,(p_1\.p_2)^4\right) \quad, \nonumber \\
\\
&C_{11}= \omega (B_{11}^1) \!+\!2 \omega^2 \left[ 2B_{001}^1\!+\!(p_1^2\!-\!p_1\.p_2) B_{111}^1 \right] \!+\!4\omega^3 \big[ 2 B_{0000}^1
\!+\!\{ 4(p_1^2\!-\!p_1\.p_2)\!+\!(p_1\!-\!p_2)^2 \} B_{0011}^1 \nonumber \\
&\,\,+(p_1^2-p_1\.p_2)^2 B_{1111}^1 - 2B_{0000}^2 \big] +8\omega^4 \big[ \{6 (p_1^2-p_1\.p_2)
+6(p_1-p_2)^2\} B_{00001}^1 \nonumber \\
&\,\,+\{ 6(p_1^2-p_1\.p_2)^2+3(p_1-p_2)^2(p_1^2-p_1\.p_2)\} B_{00111}^1+(p_1^2-p_1\.p_2)^3 B_{11111}^1 \nonumber \\
&\,\,+6 (p_2^2-p_1\.p_2)B_{00001}^2 \big]
+\mathcal{O}\left((p_1^2)^4,(p_1^2)^4,(p_1\.p_2)^4\right) \quad, \nonumber
\end{align}
\begin{align}\label{Cexpr2}
&C_{12}=C_{21}= -2 \omega^2 \left[ B_{001}^1+B_{001}^2\right] -8\omega^3 \big[ B_{0000}^1- B_{0000}^2 +(p_1^2-p_1\.p_2) B_{0011}^1 -(p_2^2-p_1\.p_2) B_{0011}^2 \big] \nonumber \\
&\,\,-8\omega^4 \big[ \{6 (p_1^2-p_1\.p_2)+3 (p_1-p_2)^2\} B_{00001}^1 +3 (p_1^2-p_1\.p_2)^2  B_{00111}^1 \nonumber \\
&\,\,+\{6 (p_2^2-p_1\.p_2)+3(p_1-p_2)^2\} B_{00001}^2 +3(p_2^2-p_1\.p_2)^2 B_{00111}^2 \big]
+\mathcal{O}\left((p_1^2)^4,(p_1^2)^4,(p_1\.p_2)^4\right) \quad, \nonumber \\
\\
&C_{22 }= \!-\!\omega (B_{11}^2) \!+\!2 \omega^2 \left[ 2B_{001}^2\!+\!(p_2^2-p_1\.p_2) B_{111}^2\right]\!+\!4\omega^3 \big[  2B_{0000}^1
\!-\!\{ 4(p_2^2\!-\!p_1\.p_2)\!+\!(p_1\!-\!p_2)^2 \} B_{0011}^2 \nonumber \\
&\,\,-(p_2^2-p_1\.p_2)^2 B_{1111}^2 - 2B_{0000}^2 \big] +8\omega^4 \big[ \{6 (p_2^2-p_1\.p_2)+6(p_1-p_2)^2\} B_{00001}^2 \nonumber \\
&\,\,+\{ 6(p_2^2-p_1\.p_2)^2+3(p_1-p_2)^2(p_2^2-p_1\.p_2)\} B_{00111}^2+(p_2^2-p_1\.p_2)^3 B_{11111}^2 \nonumber \\
&\,\,+6 (p_1^2-p_1\.p_2) B_{00001}^1\big]
+\mathcal{O}\left((p_1^2)^4,(p_1^2)^4,(p_1\.p_2)^4\right) \quad.
\end{align}

Even though four--point functions are more complicated, we can apply the same technique:
\begin{align*}
\A_1=\frac{1}{(m_1^2-m_2^2)(m_1^2-m_3^2)}\,\,,&\quad \A_2=\frac{1}{(m_2^2-m_1^2)(m_2^2-m_3^2)}\,\,, \quad \A_3=\frac{1}{(m_3^2-m_1^2)(m_3^2-m_2^2)}\,\,,
\end{align*}
\begin{align}
{}^4\!\!\F_{1,(0)}&= \frac{\A_1}{(1+\chi_{12})(1+\chi_{13})}   \,\,,\nonumber \\
{}^4\!\!\F_{1,(1)}^\mu&= -\left(\frac{2\A_1}{(1+\chi_{12})^2(1+\chi_{13})}\right) \P_{12}^\mu
-\left(\frac{2\A_1}{(1+\chi_{12})(1+\chi_{13})^2}\right)\P_{13}^\mu  \,\,,\nonumber \\
{}^4\!\!\F_{1,(2)}^{\mu\nu}&= \left(\frac{2^2 \A_1}{(1+\chi_{12})^3(1+\chi_{13})}\right) \P_{12}^\mu \P_{12}^\nu
+\left(\frac{2\A_1}{(1+\chi_{12})^2(1+\chi_{13})^2}\right) \P_{12}^\mu \P_{13}^\nu \,\,,\nonumber \\
&\quad +\left(\frac{2\A_1}{(1+\chi_{12})^2(1+\chi_{13})^2}\right) \P_{13}^\mu \P_{12}^\nu
+\left(\frac{2^2 \A_1}{(1+\chi_{12})(1+\chi_{13})^3}\right)\P_{13}^\mu \P_{13}^\nu \,\,,\nonumber \\
\cdots \,\, .
%\F_1^{4,(n)}&=\frac{2^n A_1}{n!} \sum_{i_1,\cdots,i_n=1}^{n} \left( \partial_{\chi_{12}}^{2n-r-1}\partial_{\chi_{13}}^{r}\frac{1}{(1+\chi_{12})(1+\chi_{13})}\right)
%\P_{12}^{\mu_1\cdots \mu_{2^n-r-1}} \P_{13}^{\nu_1 \cdots \nu_{r}}, \quad n\ge 1.
\end{align}
The other $\F$'s can be obtained from the permutation of $1,2$ and $3$ as follows:
\begin{align*}
{}^4\F_{2,(n)}&=\left.{}^4\F_{1,(n)}\right|_{1 \rightarrow 2,\,\, 2 \rightarrow 1} \,\,, \qquad
{}^4\F_{3,(n)}=\left.{}^4\F_{1,(n)}\right|_{1 \rightarrow 3, \,\,2 \rightarrow 1,\,\,3 \rightarrow 2} \,\,.
\end{align*}
Explicit expressions for four--point integrals are shown in Appendix~\ref{ap2}. Note that $|\chi_{ij}|<1$ must be satisfied.

\begin{figure}
\includegraphics[width=0.48\textwidth]{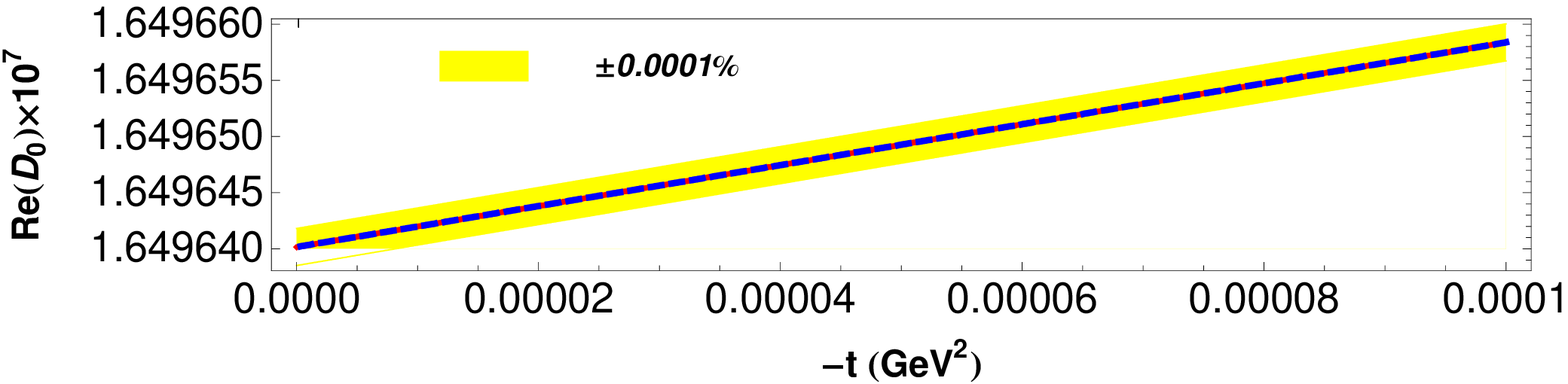}%
\includegraphics[width=0.48\textwidth]{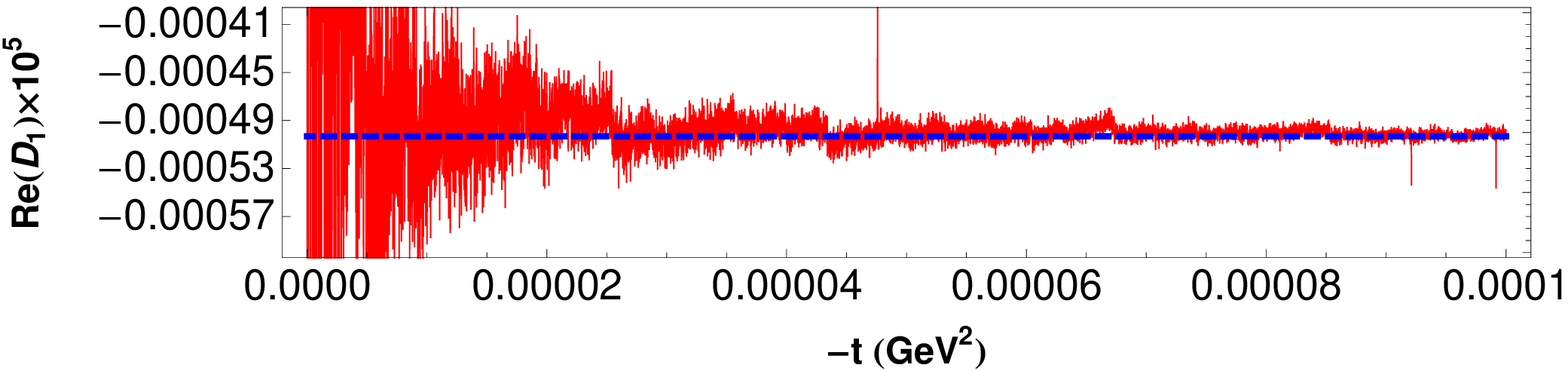}%

\includegraphics[width=0.48\textwidth]{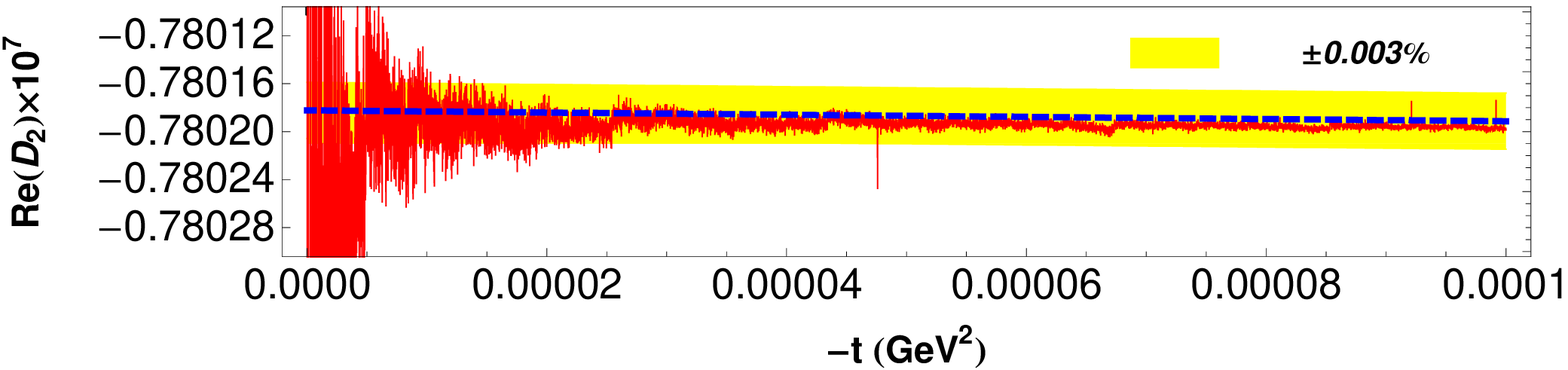}%
\includegraphics[width=0.48\textwidth]{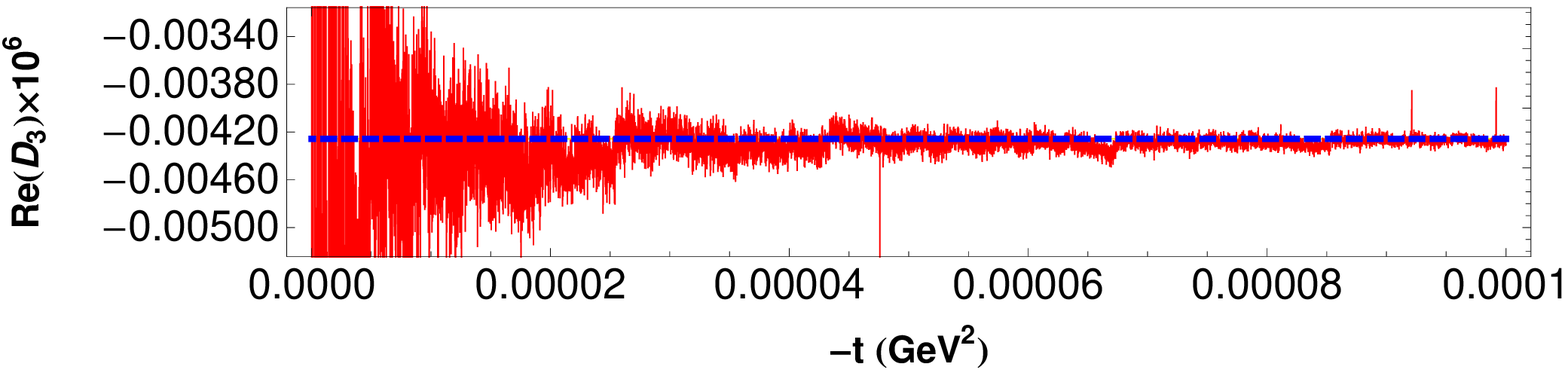}%

\includegraphics[width=0.48\textwidth]{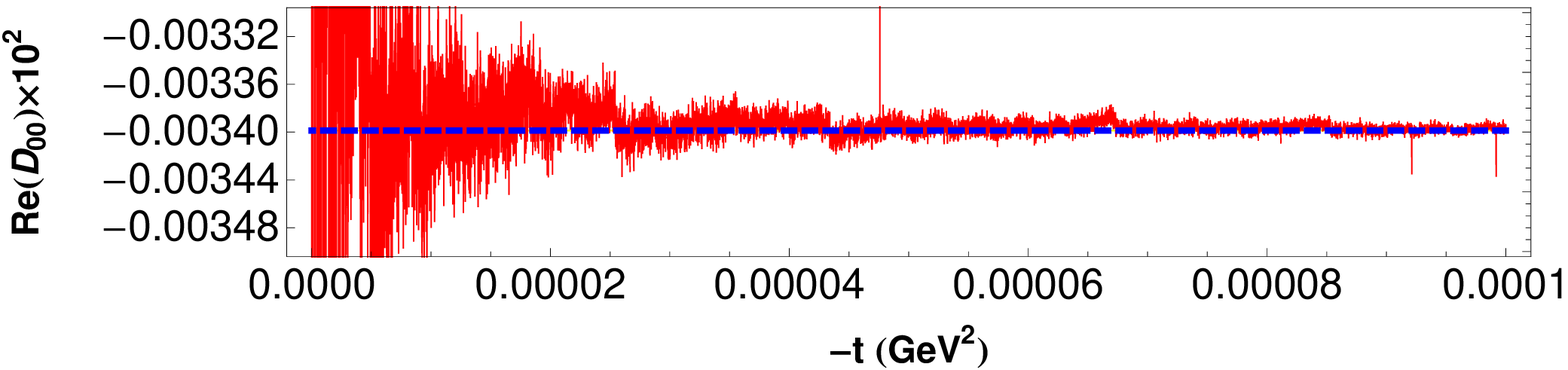}%
\includegraphics[width=0.48\textwidth]{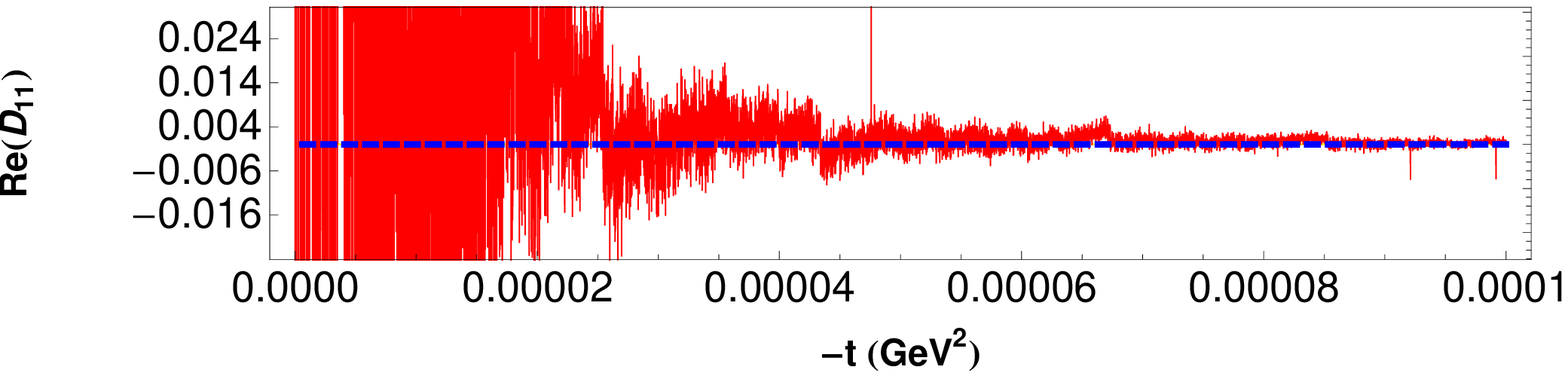}%

\includegraphics[width=0.48\textwidth]{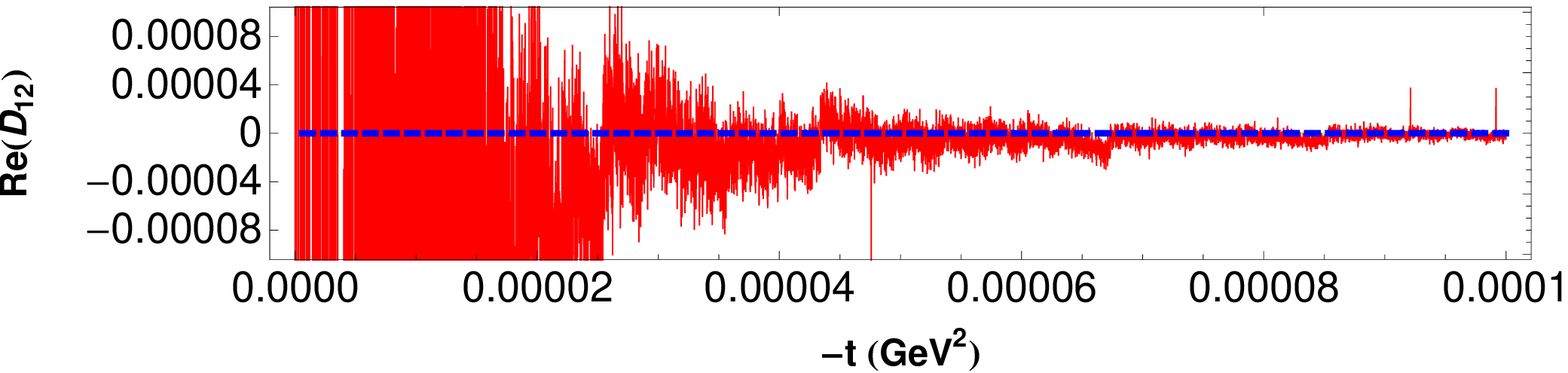}%
\includegraphics[width=0.48\textwidth]{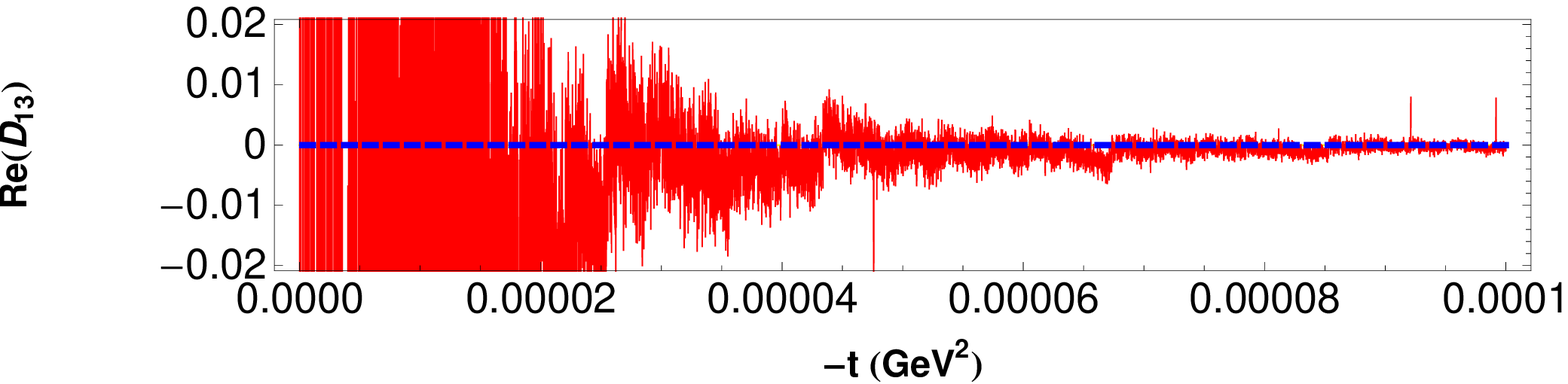}%

\includegraphics[width=0.48\textwidth]{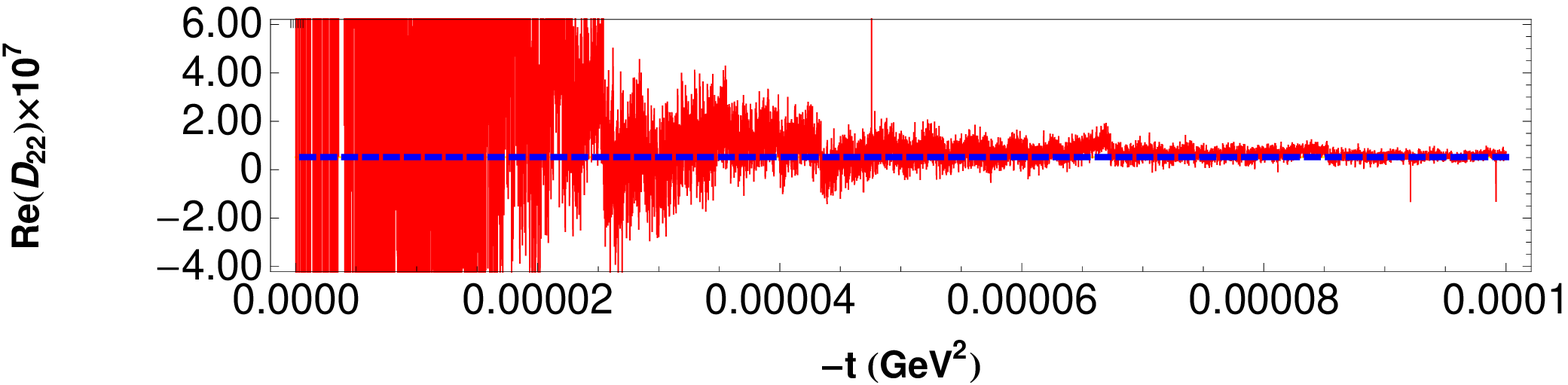}%
\includegraphics[width=0.48\textwidth]{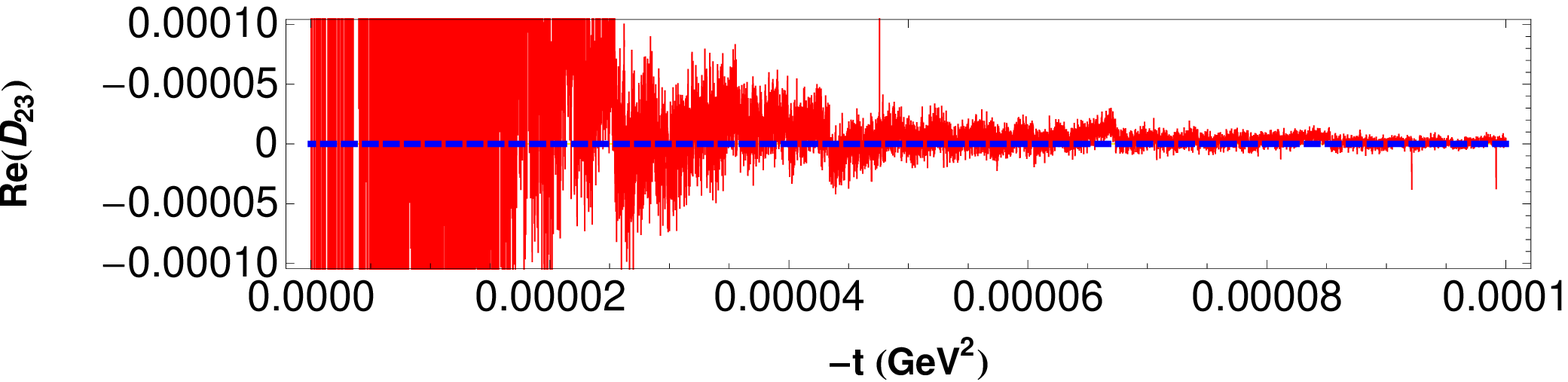}%

\includegraphics[width=0.48\textwidth]{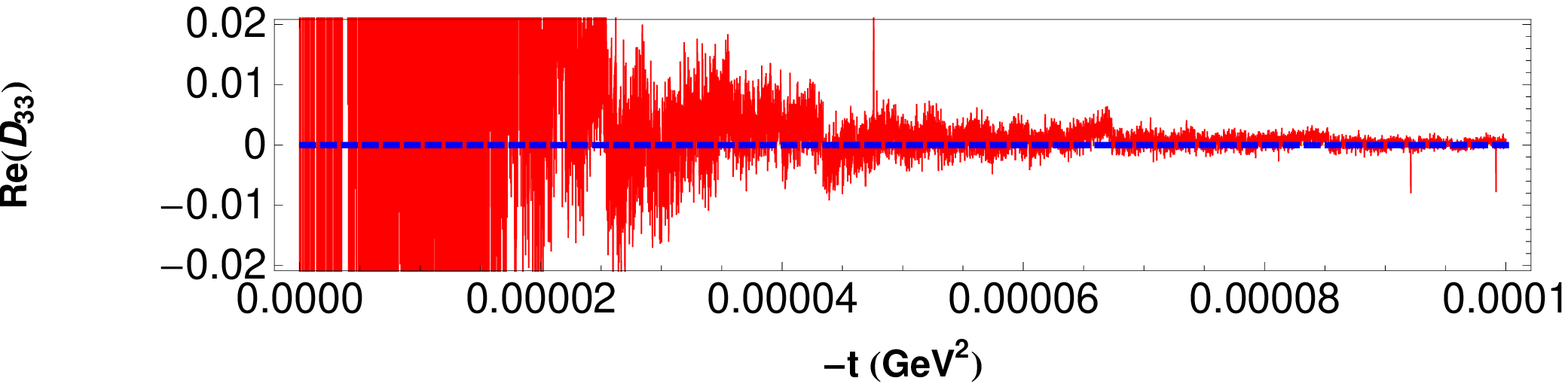}%
\caption{\label{figDsm}Numerical safety of four-point functions : $D_{\cdots}[p_1^2, p_2^2, p_3^2, (p_1-p_2)^2, (p_2-p_3)^2, (p_1-p_3)^2, m_1^2, m_2^2, m_3^2, m_4^2]$$=$$D_{\cdots}[0, u, 0, 0, t, m_d^2, M_W^2, 0, M_Z^2]$ with $m_d=0.066\,\,GeV$. The Mandelstam variable, t varies from $-0.0001$ to $0$, and $s=1\,\,GeV^2$.
The conventional Passarino-Veltman reduction shows numerical instability in small kinematic region (red curve),
while blue curve which is generated from the method in Sec.~\ref{sec3}(Eq.~\eqref{Dexpr}) is numerically stable. The Feynman diagram is shown in Fig.~\ref{fdCD}. Red curves are generated using {\it LoopTools}~\cite{Hahn:2000jm}.}
\end{figure}

The numerical computations of the $C$ and $D$ functions are shown in both Fig.~\ref{figCsm} and Fig.~\ref{figDsm}. Although $C_0$ and $D_0$ don't have any numerical problem, the approximation of small kinematic variables is in accord with the analytical calculation.

\section{\label{sec4} Six and more point  functions, F,G,H....: Alternative Reduction}

\begin{figure}
\includegraphics[width=0.48\textwidth]{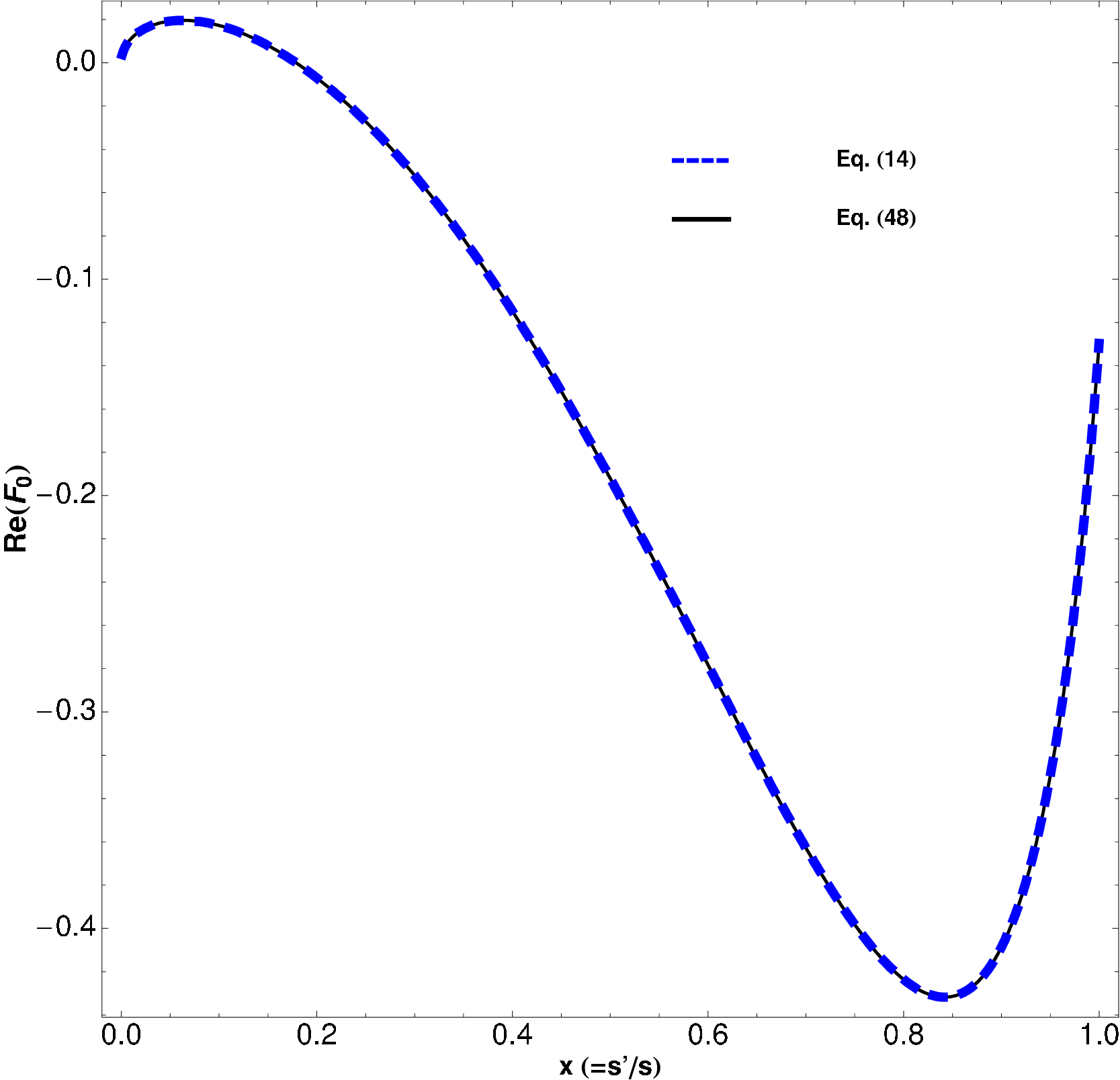}
\caption{\label{figF}Comparison of six--point scalar integrals using Eq.~\eqref{master1} (or Ref.~\cite{Denner:2005nn}) and Eq.~\eqref{master6}.
The two curves are identical. This not only shows the validity of \eqref{master6} but also improves numerical efficiency. This computation is based on the Feynman diagram in Fig.~\ref{fdEF}. }
\end{figure}

Here, we suggest an alternative reduction formula for the $n$-point integrals that are always decomposed to six $(n-1)$-point functions.
Without loss of generality, $p_1,\,p_2,\,p_3$ and $p_4$ are supposed to be linearly independent. A vector $p_\ell$, where $\ell$ is one of the integers of $5 \le \ell \le n-1$, can be expressed as:
\begin{align*}
p_{\ell}=\sum_{i=1}^4 x_{i} p_i.
\end{align*}
The $x_i$'s can be found using the $4\times4$ Gram matrix, $Z_{ij}=2p_i \. p_j$ with $i,j=1,2,3,4$:
\begin{align}\label{r6x}
x_i=2 \sum_{j=1}^4 \left(Z^{\!-1}\right)_{ij} p_\ell \.p_j \quad.
\end{align}
Therefore, a system of equations, Eq.~\eqref{cons-1},  Eq.~\eqref{cons-2},and  Eq.~\eqref{cons-3} can be simplified as:
\begin{align*}
a_1+a_\ell x_{1}&=0\,\,, \quad a_2+a_\ell x_{2}=0\,\,, \\ a_3+a_\ell x_{3}&=0\,\,, \quad a_4+a_\ell x_{4}=0\,\,, \\
a_0+a_1+\cdots + a_{n-1}&=0 \,\,,\quad \sum_{i=0}^{n-1} (p_i^2-m_i^2)a_i=1 \,\,.
\end{align*}
Hence, all coefficients $a_i$ are easily determined as:
\begin{align}\label{r6a}
a_i&=\frac{-x_i}{f_\ell-\sum_j^4 f_j x_j}\,\,, \,\,(i=1,2,3,4) \nonumber \\
a_\ell&=\frac{1}{f_\ell-\sum_j^4 f_j x_j}\,\,, \nonumber \\
a_0&=\frac{1-x_1-x_2-x_3-x_4}{f_\ell-\sum_j^4 f_j x_j}\,\, ,
\end{align}
where $f_\ell=p_\ell^2-m_\ell^2+m_0^2$, and $\ell$ is one of the integers of $5 \le \ell \le n-1$.
With the above $a_i$'s, the identity for the reduction is:
\begin{align*}
\frac{1}{N_0\cdots N_{n-1}}=\frac{a_\ell N_\ell}{N_0\cdots N_{n-1}}+\sum_{i=0}^4 \frac{a_i N_i}{N_0\cdots N_{n-1}}\quad.
\end{align*}
Finally, an alternative reduction formula is:
\begin{align}\label{master6}
T^n=a_\ell T^{n-1}(\ell)+\sum_{i=0}^4 a_i T^{n-1}(i), \,\,(n\ge 6),
\end{align}
where $\ell$ is one of the integers $\{5,\cdots,n-1\}$. The coefficients $a_i$ are determined from the $x_i$ which can be obtained from the Gram matrix (Eq.~\eqref{r6x} and Eq.~\eqref{r6a}). Although the alternative reduction of Eq.~\eqref{master6} uses the Gram matrix, there is no problem of the vanishing Gram determinant because $p_1,\,p_2,\,p_3$ and $p_4$ are linearly independent. Also, the case of a small Gram determinant is already derived in Sec.~\ref{sec3}.

According to the reduction formula of Eq.~\eqref{master1}, we should compute the inverse of $(n+1)\times(n+1)$ Caley matrix.
Here, the alternative reduction doesn't need any computation of a high--dimensional Caley matrix regardless of the value of $n$.
In the previous method, $n$--point integral reduces to a sum of $n$ number of $(n-1)$-point integrals. However, the alternative way gives always $6$ number of $(n-1)$-point integrals which improves the efficiency of numerical computation. Vector and tensor coefficient functions can be obtained using the same technique in Sec.~\ref{sec2}.
\section{\label{sec5}Conclusion}
\begin{figure}
\includegraphics[width=0.5\textwidth]{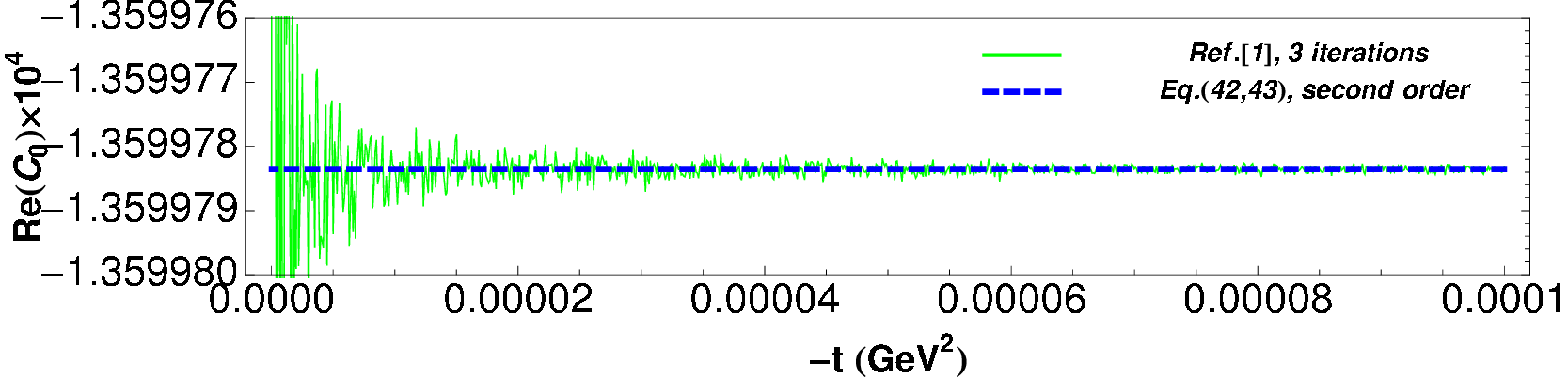}
\includegraphics[width=0.5\textwidth]{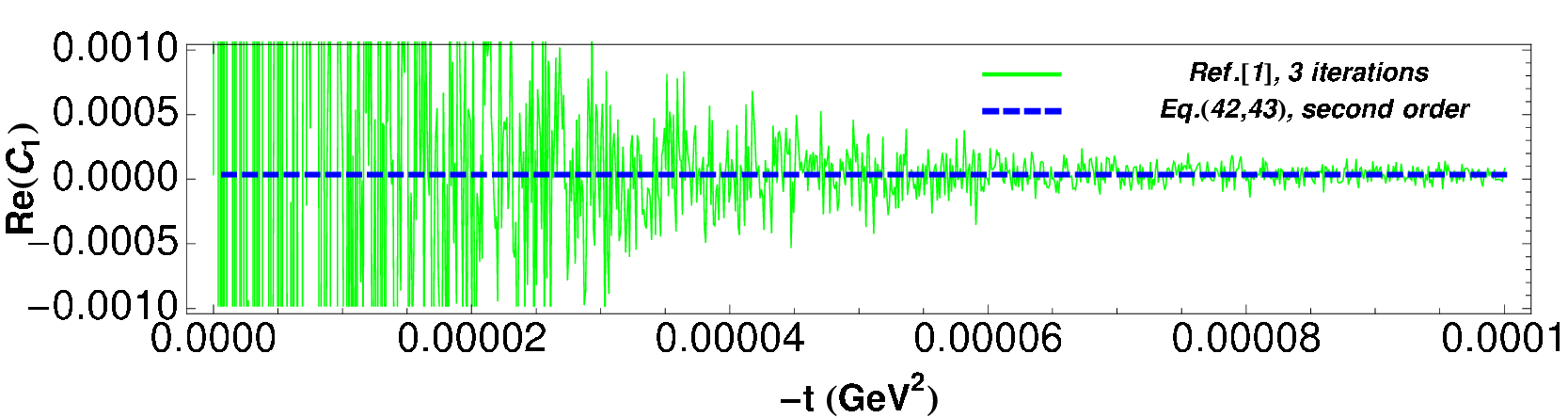}%
\includegraphics[width=0.5\textwidth]{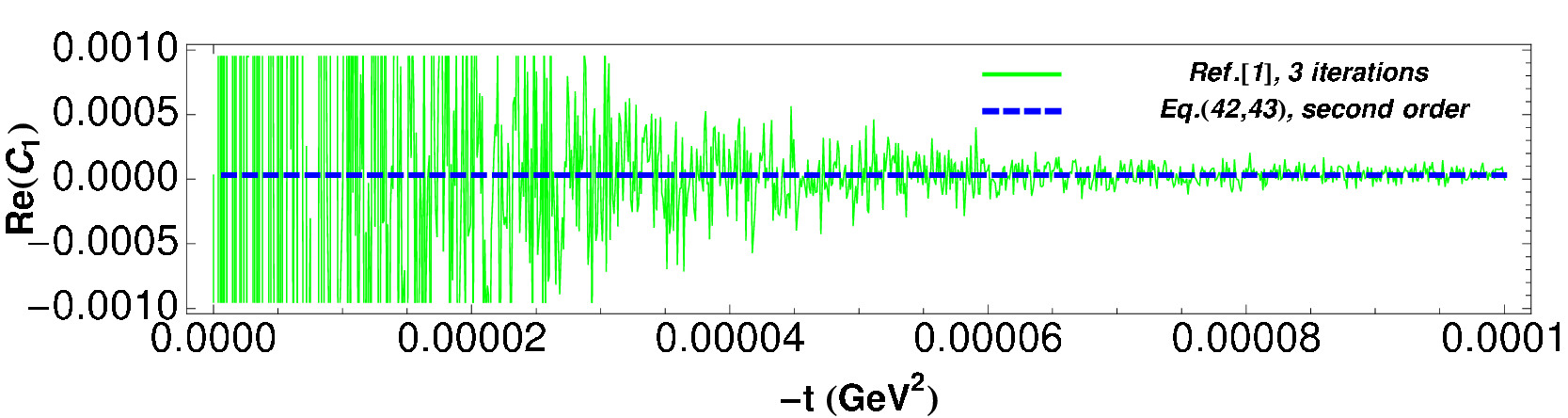}%
\caption{\label{comDD} Comparison between the method in Sec.\ref{sec3} (Eq.~\eqref{Cexpr1},~\eqref{Cexpr2}) and the method described in Ref.~\cite{Denner:2005nn}. We computed $Re[C_{0,1,2}]$ up to second order using Eq.~\eqref{Cexpr1},~\eqref{Cexpr2} (dashed line), while for the method in Ref.~\cite{Denner:2005nn}, three iterations used (solid curve), in which each iteration corresponds to the order of momentum squared. In order for the method in Ref.~\cite{Denner:2005nn} to show reasonable stability, we should perform much more iterations.}
\end{figure}

We formulate a reduction for $n$--point integrals with a new formalism.
In the region of small kinematic variables, which yield a small Gram determinant, it is well known that NLO calculation suffers from numerical instability when we use the conventional Passarino-Veltman reduction. The analytical expression for that region not only improves the numerical efficiency, but also fixes the numerical instability so long as the condition $|\chi_{ij}|<1$ is satisfied. In Fig.~\ref{figCsm} and Fig.~\ref{figDsm}, we showed the numerical stability of three--point and four--point integrals. The \emph{numerical} method described in Ref.~\cite{Denner:2005nn} fixes the numerical instability as well. In which each order of a small Gram determinant is computed iteration by iteration. However, the efficiency of the numerical computation is much improved as shown in Fig.~\ref{comDD} by the \emph{analytical} method which provides faster convergence of the series.

The reduction formula for the one--loop $n$--point integrals with $n\ge6$, are derived alternatively in Sec.~\ref{sec3} (Eq.~\eqref{master6}), where we reduced the number of sub--integrals. As an example,  if we use Eq.~\eqref{master1} or the method suggested in Ref.~\cite{Denner:2005nn}, then the eight--point scalar integral is decomposed in 8 seven--point scalar integrals. However, with the newly derived reduction formula, Eq.~\eqref{master6}, we needs only 6 seven--point scalar integrals. Additionally, because we don't use the Caley matrix, we can reduce the uncertainty which enters when computing the inverse of a high--dimensional Cayley matrix, thereby saving CPU time. Also, we can avoid the case of the vanishing Caley determinant.

\begin{acknowledgments}
I am grateful to Doreen Wackeroth for the valuable discussions. Also, I would like to thank Fred Olness and Pavel Nadolsky for advice. The work of K. P. is supported by the U.S. Department of Energy under grant DE-FG02-04ER41299.
\end{acknowledgments}

\appendix
\section{\label{ap1}Two-point integrals: $B$-functions}
For the $B$-function expansion, we show the explicit expressions of the $B$-functions and their derivative in this Appendix.
We begin by defining the following integrals:
\begin{align} \label{ls}
L_{mn}(a;\alpha,\beta;x_0,x_1)&=\int_0^1 dx \frac{\log\left(a(x-\alpha)(x-\beta)\right)}{(x-x_0)^m(x-x_1)^n } \quad,\nonumber \\
S_{k\ell}^{mn}(a;\alpha,\beta;x_0,x_1)&=\int_0^1 dx \frac{a (x-\alpha)^m(x-\beta)^n }{(x-x_0)^k(x-x_1)^\ell}\quad,
\end{align}
where $L_{00}(a;\alpha,\beta)$ is:
\begin{align}
&L_{00}(a;\alpha,\beta)=\int_0^1 dx \log\left( a(x-\alpha)(x-\beta) \right)=\nonumber \\
&\left\{ \begin{matrix}
-2 - \alpha \log(1 - \alpha) + \alpha \log(-\alpha) - \beta \log(1 - \beta) & &\\ 
+\log\left(a (1 - \alpha)(1 - \beta)\right) + \beta \log(-\beta), & & \alpha,\beta \ne 0,1 \\ \\
-2 - \beta \log(1 - \beta) + \beta \log(-\beta) +
\log\left(a(1 - \beta)\right),& & \alpha=0 \,\& \,\beta\ne0,1 \\ \\
-2 + \log(a),& &\alpha=0 \,\& \,\beta=0 \\ \\
-2 + \log(-a),& &\alpha=0 \,\& \,\beta=1
\end{matrix} \right. \label{s12}
\quad ,
\end{align}
which is directly connected to the $B_0$ function; $B_0(p_1^2,m_0^2, m_1^2)=-L_{00}(a;\alpha,\beta)$ with $a=p_1^2$. The $\alpha$, and $\beta$ should satisfy the following quadratic equation:
\begin{align*}
p_1^2 x^2+(m_1^2-m_0^2-p_1^2)x + m_0^2-i\epsilon=0 \quad.
\end{align*}
Because $S_{k\ell}^{mn}$ has numerical instability in high $x$, let us introduce a numerically safe function $f_n (x)$:
\begin{equation*}
f_n(x)=\left\{ 
\begin{array}{cr} \displaystyle \frac{x}{n-1}+\frac{x^2}{n-2}+ \cdots +\frac{x^{n-1}}{1}+ x^n \log\left(1-\tfrac{1}{x}\right) & {\rm ,\,\, for}\,\, x < 10^{7-n}\,\,.\\ \displaystyle
-\frac{1}{n}-\sum_{i=1}^\infty \frac{1}{x^i(n+i)} & {\rm ,\,\,for} \,\, x \ge 10^{7-n} \,\,.
\end{array} \right. \quad, 
\end{equation*}
where the number $10^{7-n}$ has been chosen for the computations shown in Fig.~\ref{figCsm} and~\ref{figDsm}. For the expression of the $B$-functions, we need the followings with the condition of $n>-1$:
\begin{align*}
S_{m,m}^{n,0}(\tfrac{1}{a^m};0;\alpha,\beta)&= \frac{F_1\left[ n+1,m,m,n+2,\tfrac{1}{\alpha},\tfrac{1}{\beta} \right]}
{a^m \alpha^m \beta^m (n+1)}  \,\,,\\
S_{m,m}^{n,1}(\tfrac{1}{a^m};0,\tfrac{\alpha+\beta}{2};\alpha,\beta)&=\frac{1}{a^m \alpha^m \beta^m (n+1)}\Bigg(
(\beta-\alpha)F_1\left[ n+1,m,m,n+2,\tfrac{1}{\alpha},\tfrac{1}{\beta} \right] \\
&\quad+2(n+1)F_1\left[ n+2,m,m,n+3,\tfrac{1}{\alpha},\tfrac{1}{\beta} \right]\Bigg) \quad.
\end{align*}
where Appell hypergeometric function $F_1$ is introduced, which can be expressed in terms of $f_n$ using the identities of Ref.~\cite{Murley:2008}. As examples:
\begin{align*}
S_{1,1}^{n,0}(\tfrac{1}{a};0;\alpha,\beta)&= \frac{\big(f_n(\alpha)-f_n(\beta)\big)}{a (\alpha-\beta) } \,\,,\\
S_{1,1}^{n,1}(\tfrac{1}{a};0,\tfrac{\alpha+\beta}{2};\alpha,\beta)&=\frac{2}{an}+\frac{1}{a}\left(f_n(\alpha)+f_n(\beta)\right)\,\,.
\end{align*}
Any order of derivatives with respect to $m_1^2$ can be obtained from the following recursion relations:
\begin{align*}
\partial_1 S_{m,m}^{n,0}(\tfrac{1}{a^m};0;\alpha,\beta)&=-m S_{m+1,m+1}^{n+1,0}(\tfrac{1}{a^m};0;\alpha,\beta)\,\,,\\
\partial_1 S_{m,m}^{n,1}(\tfrac{1}{a^m};0,\tfrac{\alpha+\beta}{2};\alpha,\beta)&=\frac{1}{2} S_{m,m}^{n,0}(\tfrac{1}{a^{m+1}};0;\alpha,\beta)-m S_{m+1,m+1}^{n+1,1}(\tfrac{1}{a^{m+1}};0,\tfrac{\alpha+\beta}{2};\alpha,\beta)\,\,,
\end{align*}
where $\partial_1=\tfrac{\partial}{\partial m_1^2}$.

The definition of $L_{00}$ gives the first derivative of the two-point scalar integral:
\begin{align*}
\partial_1 B_0&=-\int_0^1dx \partial_1 \log(a (x-\alpha)(x-\beta))
=-\int_0^1 dx\frac{x}{a (x-\alpha)(x-\beta)}
=-S_{1,1}^{1,0}(\tfrac{1}{a};0;\alpha,\beta)\,\,.
\end{align*}

Although the two-point vector and tensor coefficient functions are already shown in Ref.~\cite{Denner:2005nn}, we briefly derive an explicit form for them:
\begin{align}\label{bn}
B_{\underbrace{\scriptscriptstyle{1\cdots1}}_{n}}=\int_0^1dx (-x)^n\left[\Delta+\log\mu^2-\log(a (x-\alpha)(x-\beta))\right] \,\,,
\end{align}
where UV--divergence term $\Delta$ is defined as $\Delta=\tfrac{2}{4-D}-\gamma_E+\log(4\pi)$, and $\gamma_E$ is Euler's constant. For simplicity, the constant term,   $\Delta+\log\mu^2$ will be omitted in the following. After integration by parts, Eq.~\eqref{bn} simplifies as:
\begin{align*}
B_{\underbrace{\scriptscriptstyle{1\cdots1}}_{n}}=-\frac{(-1)^n}{n+1}\Big( \log(a(1-\alpha)(1-\beta))- 2aS_{1,1}^{n+1,1}(\tfrac{1}{a};0,\tfrac{\alpha+\beta}{2};\alpha,\beta) \Big) \,\,.
\end{align*}

Now, we summarize the two-point functions and their first derivatives when $p_1^2 \ne 0$:
\begin{align*}
B_{\underbrace{\scriptscriptstyle{1\cdots1}}_{n}}&=\frac{(-1)^{n+1}}{n+1}\Big( \log(a(1-\alpha)(1-\beta))- 2 aS_{1,1}^{n+1,1}(\tfrac{1}{a};0,\tfrac{\alpha+\beta}{2};\alpha,\beta) \Big), \\
B_0&=-L_{00}(a;\alpha,\beta) \\
B_{00}&=\frac{1}{6}\Big(m_1^2(1-\log(m_1^2))+(p_1^2-m_1^2+m_0^2)B_1  + 2m_0^2 B_0+m_0^2+m_1^2-\tfrac{1}{3}p_1^2 \Big), \\
B_{001}&=\frac{1}{8}\Big(-m_1^2(1-\log(m_1^2))+(p_1^2-m_1^2+m_0^2)B_{11}  + 2m_0^2 B_1-\tfrac{1}{6}(2m_0^2+4m_1^2-p_1^2) \Big), \\
\cdots,
\end{align*}
\begin{align*}
B_{\underbrace{\scriptscriptstyle{1\cdots1}}_{n}}'&=\frac{(-1)^{n+1}}{n+1}\Big( \frac{1}{a(1-\alpha)(1-\beta)}-  aS_{1,1}^{n+1,0}(\tfrac{1}{a^{2}};0;\alpha,\beta)+2a S_{2,2}^{n+2,1}(\tfrac{1}{a^{2}};0,\tfrac{\alpha+\beta}{2};\alpha,\beta) \Big),\\
B_0'&=-S_{11}^{10}(\tfrac{1}{a};0;\alpha,\beta), \\
B_{00}'&=\frac{1}{6}\Big(1-\log(m_1^2)-B_1'+(p_1^2-m_1^2+m_0^2)B_1'+2m_0^2B_0'\Big), \\
B_{001}'&=\frac{1}{8}\Big( \log(m_1^2)-B_{11}+(p_1^2-m_1^2+m_0^2)B_{11}'+2m_0^2B_1'-\frac{2}{3}\Big), \\
\cdots.
\end{align*}
When $p_1^2 = 0$:
\begin{align*}
B_0&=\frac{1}{m_0^2-m_1^2}\Big(m_0^2-m_1^2-m_0^2 \log(m_0^2)+m_1^2\log(m_1^2)\Big), \\
B_1&=\frac{1}{4(m_0^2-m_1^2)^2}\Big(-3 m_0^4+4m_0^2 m_1^2-m_1^4+2 m_0^4 \log(m_0^2)-2 m_1^2(2m_0^2-m_1^2)\log(m_1^2)\Big), \\
B_{00}&=\frac{1}{8(m_0^2-m_1^2)}\Big(3 m_0^4-3 m_1^4 -2 m_0^4 \log(m_0^2)+2 m_1^4\log(m_1^2)\Big), \\
B_{11}&=-\frac{1}{18(m_0^2-m_1^2)^3}\Big( 6 m_0^6 \log(m_0^2)-6 m_1^6\log(m_1^2) \\
 &\quad +(m_0^2-m_1^2)(-11m_0^4+7m_0^2m_1^2-2 m_1^4+6(m_0^2-m_1^2)^2 \log(m_1^2)\Big), \\
B_{001}&=\frac{1}{72(m_0^2-m_1^2)^2}\Big(-11m_0^6+27 m_0^2 m_1^4-16 m_1^6-3 m_0^6\log(m_0^2)+9m_0^6\log(m_0^2)+3m_0^6 \log(m_1^2)\Big), \\
B_{111}&=\frac{1}{48(m_0^2-m_1^2)^4}\Big(12 m_0^8\log(m_0^2)-12m_0^8\log(m_1^2)\\
&\quad +(m_0^2-m_1^2)(-25 m_0^6+23m_0^4m_1^2-13m_0^2m_1^4+3m_1^6+12(m_0^2-m_1^2)^3\log(m_1^2) \Big).\\
\cdots.
\end{align*}

For special cases, such as $m_0=0$ or $m_0=m_1$, we have to start from the expression of Eq.~\eqref{ls}.

\section{\label{ap2}$D$-functions for the region of small kinematic variables}
For the four-point functions described in Sec.~\ref{sec3} for the general mass case, the explicit expressions in small Gram determinant region are:
\begin{align*}
D_0&=a_1 B_0^1 \!+ \!a_2 B_0^2 \!+ \!a_3 B_0^3\!+\!b_{11} ( p_1^2 B_1^1 \!- \!p_1\.p_2 B_1^1 )
\!+\!b_{12} ( p_2^2 B_1^2 \!- \!p_1\.p_2 B_1^2 )\!+\!b_{13} ( p_3^2 B_1^3 \!- \!p_1\.p_3 B_1^3 )\\
&+b_{21} ( p_1^2 B_1^1 - p_1\.p_3 B_1^1 )+b_{22} ( p_2^2 B_1^2 - p_2\.p_3 B_1^2 )
+b_{23} ( p_3^2 B_1^3 - p_2\.p_3 B_1^3 )\\
&+c_{11} \left( (p_1-p_2)^2 B_{00}^1 + (p_1^2-p_1\.p_2)^2 B_{11}^1 \right)
+c_{12} \left( (p_1-p_2)^2 B_{00}^2 + (p_2^2-p_1\.p_2)^2 B_{11}^2 \right)\\
&+c_{13} \left( (p_1-p_3)^2 B_{00}^3 + (p_3^2-p_1\.p_3)^2 B_{11}^3 \right)\\
&+2c_{21}\big((p_1^2-p_1\.p_2-p_1\.p_3+p_2\.p_3)B_{00}^1+(p_1^2-p_1\.p_2)(p_1^2-p_1\.p_3)B_{11}^1\big)\\
&
+2c_{22}\left((p_2^2-p_1\.p_2+p_1\.p_3-p_2\.p_3)B_{00}^2+(p_2^2-p_1\.p_2)(p_2^2-p_2.p_3)B_{11}^2\right)\\
&
+2c_{23}\left((p_3^2+p_1\.p_2-p_1\.p_3-p_2\.p_3) B_{00}^3+(p_3^2-p_1\.p_3)(p_3^2-p_2\.p_3)B_{11}^3\right)\\
&
+c_{31} \left( (p_1-p_3)^2 B_{00}^1 + (p_1^2-p_1\.p_3)^2 B_{11}^1 \right)
+c_{32} \left( (p_2-p_3)^2 B_{00}^2 + (p_2^2-p_2\.p_3)^2 B_{11}^2 \right)\\
&
+c_{33} \left( (p_2-p_3)^2 B_{00}^3 + (p_3^2-p_2\.p_3)^2 B_{11}^3 \right) \,\,,
\end{align*}
\begin{align*}
D_1&=a_1 B_1^1+b_{11}\left(B_{00}^1+(p_1^2-p_1\.p_2)B_{11}^1\right)-b_{12}B_{00}^2-b_{13}B_{00}^3
+b_{21} \left(B_{00}^1 + (p_1^2 - p_1\.p_3) B_{11}^1\right)\\
&
\!+\!c_{11} \!\left( (3 p_1^2 \!- \!4 p_1\.p_2 \!+ \!p_2^2) B_{001}^1 \!\!+ \!(p_1^2 \!- \!p_1\.p_2)^2 B_{111}^1\right)
\!+\!2c_{12}(p_1\.p_2 \!- \!p_2^2) B_{001}^2\!\!+\!2c_{13}(p_1\.p_3 \!- \!p_3^2) B_{001}^3\\
&
+2c_{21} \left( (3 p_1^2 - 2 (p_1\.p_2 + p_1\.p_3) + p_2\.p_3) B_{001}^1 + (p_1^2 - p_1\.p_2) (p_1^2 - p_1\.p_3) B_{111}^1 \right)\\
&
\!-\!2c_{22} (p_2^2 \!- \!p_2\.p_3) B_{001}^2
\!\!-\!2c_{23} (p_3^2 \!- \!p_2\.p_3) B_{001}^3
\!\!+\!c_{31} \left( (3 p_1^2 \!- \!4 p_1\.p_3 \!+ \!p_3^2) B_{001}^1 \!+ (\!p_1^2 \!- \!p_1\.p_3)^2 B_{111}^1\right) \,\,,
\end{align*}
\begin{align*}
D_2&=a_2 B_1^2-b_{11}B_{00}^1+b_{12} \left(B_{00}^2 + (p_2^2-p_1\.p_2) B_{11}^2\right)
+b_{22} \left( B_{00}^2 + (p_2^2 - p_2\.p_3) B_{11}^2\right)-b_{23} B_{00}^3\\
&
\!-\!2c_{11}(p_1^2 \!-\! p_1\.p_2) B_{001}^1
\!\!+\!c_{12}\left((p_1^2 \!-\! 4 p_1\.p_2\! +\! 3 p_2^2) B_{001}^2\!\! +\! (p_2^2 \!-\! p_1\.p_2 )^2 B_{111}^2\right)
\!-\!2c_{21}(p_1^2\! -\! p_1\.p_3) B_{001}^1\\
&
+2c_{22}\left(( 3 p_2^2-2( p_1\.p_2 + p_2\.p_3)+ p_1\.p_3  ) B_{001}^2 + (p_2^2-p_1\.p_2 ) (p_2^2 - p_2\.p_3) B_{111}^2\right)\\
&
\!-\!2c_{23}(p_3^2\! -\! p_1\.p_3 ) B_{001}^3
\!\!+\!c_{32}\left((3 p_2^2 \!-\! 4 p_2\.p_3\! +\! p_3^2) B_{001}^2\!\! +\! (p_2^2 - p_2\.p_3)^2 B_{111}^2\right)
\!+\!2c_{33}(p_2\.p_3\! -\! p_3^2) B_{001}^3  \,\,,
\end{align*}
\begin{align*}
D_3&=a_3 B_1^3+b_{13}\left(B_{00}^3 + (p_3^2-p_1\.p_3) B_{11}^3\right)-b_{21}B_{00}^1-b_{22}B_{00}^2
+b_{23}\left(B_{00}^3 + (p_3^2-p_2\.p_3) B_{11}^3\right)\\
&
\!+\!c_{13}\left((p_1^2\! -\! 4 p_1\.p_3\! +\! 3 p_3^2) B_{001}^3\! +\! (p_1\.p_3\! -\! p_3^2)^2 B_{111}^3\right)
\!-\!2c_{21} (p_1^2\! -\! p_1\.p_2) B_{001}^1
\!+\!2c_{22}(p_1\.p_2 \!-\! p_2^2) B_{001}^2\\
&
+2c_{23}\left((p_1\.p_2 - 2 (p_1\.p_3 + p_2\.p_3) + 3 p_3^2) B_{001}^3 + (p_3^2-p_1\.p_3) (p_3^2-p_2\.p_3) B_{111}^3\right)\\
&
\!-\!2c_{31}(p_1^2\! -\! p_1\.p_3) B_{001}^1
\!\!+2c_{32}(p_2^2\! -\! p_2\.p_3) B_{001}^2
\!+\!c_{33}\left((p_2^2\! -\! 4 p_2\.p_3\! +\! 3 p_3^2) B_{001}^3\! +\! (p_3^2\! -\! p_2\.p_3 )^2 B_{111}^3\right)  \,\,,
\end{align*}
\begin{align*}
D_{00}&=a_1B_{00}^1\!+\!a_2B_{00}^2\!+\!a_3B_{00}^3\!+\!b_{11}(p_1^2\! -\! p_1\.p_2)B_{001}^1\!+\!b_{12}(p_2^2\!-\!p_1\.p_2)B_{001}^2
\!+\!b_{13}(p_3^2\!-\!p_1\.p_3) B_{001}^3\\
&+b_{21}(p_1^2 - p_1\.p_3) B_{001}^1+b_{22}(p_2^2 - p_2\.p_3) B_{001}^2
+b_{23}(p_3^2-p_2\.p_3) B_{001}^3\\
&
+c_{11}\left((p_1^2 - p_1\.p_2)^2 B_{0011}^1 + (p_1-p_2)^2 B_{0000}^1\right)
+c_{12}\left((p_2^2 - p_1\.p_2)^2 B_{0011}^2 + (p_1-p_2)^2 B_{0000}^2\right)\\
&
+c_{13}\left((p_3^2 - p_1\.p_3)^2 B_{0011}^3 + (p_1-p_3)^2 B_{0000}^3\right)\\
&
+2c_{21}\left( (p_1^2 - p_1\.p_2) (p_1^2 - p_1\.p_3) B_{0011}^1 +
  (p_1^2 - p_1\.p_2 - p_1\.p_3 + p_2\.p_3) B_{0000}^1\right)\\
&
+2c_{22}\left((p_2^2-p_1\.p_2) (p_2^2 - p_2\.p_3) B_{0011}^2 + ( p_2^2-p_1\.p_2 + p_1\.p_3 - p_2\.p_3) B_{0000}^2\right)\\
&
+2c_{23}\left((p_3^2 - p_1\.p_3) (p_3^2 - p_2\.p_3) B_{0011}^3 + ( p_3^2+p_1\.p_2 - p_1\.p_3 - p_2\.p_3 ) B_{0000}^3\right)\\
&
+c_{31}\left((p_1^2 - p_1\.p_3)^2 B_{0011}^1 + (p_1-p_3)^2 B_{0000}^1\right)
+c_{32}\left((p_2^2 - p_2\.p_3)^2 B_{0011}^2 + (p_2-p_3)^2 B_{0000}^2\right)\\
&
+c_{33}\left((p_3^2 - p_2\.p_3)^2 B_{0011}^3 + (p_2-p_3)^2 B_{0000}^3\right) \,\,,
\end{align*}
\begin{align*}
D_{11}&=a_1 B_{11}^1+b_{11}\left(2 B_{001}^1 + (p_1^2 - p_1\.p_2) B_{111}^1\right)
+b_{21}\left(2 B_{001}^1 + (p_1^2 - p_1\.p_3) B_{111}^1\right)\\
&
+c_{11}\left((5 p_1^2 - 6 p_1\.p_2 + p_2^2) B_{0011}^1 +
 2 B_{0000}^1 + (p_1^2 - p_1\.p_2)^2 B_{1111}^1\right)
+2c_{12}B_{0000}^2+2c_{13}B_{0000}^3\\
&
+2c_{21}\left((5 p_1^2 - 3 (p_1\.p_2 + p_1\.p_3) + p_2\.p_3) B_{0011}^1 +
 2 B_{0000}^1 + (p_1^2 - p_1\.p_2) (p_1^2 - p_1\.p_3) B_{1111}^1\right)\\
&
+c_{31}\left((5 p_1^2 - 6 p_1\.p_3 + p_3^2) B_{0011}^1 +
 2 B_{0000}^1 + (p_1^2 - p_1\.p_3)^2 B_{1111}^1\right) \,\,,
\end{align*}
\begin{align*}
D_{12}& \!=\!D_{21}\!=\!-\!b_{11} B_{001}^1\!\!-\!b_{12} B_{001}^2
\!\!-\!2 c_{11}\!\left((p_1^2\! -\! p_1\.p_2) B_{0011}^1\! +\! B_{0000}^1\right)
\!-\!2c_{12}\!\left((p_2^2\!-\!p_1\.p_2 ) B_{0011}^2\! + \! B_{0000}^2\right)\\
&
-2c_{21}\left((p_1^2 - p_1\.p_3) B_{0011}^1 + B_{0000}^1\right)
-2c_{22}\left((p_2^2 - p_2\.p_3) B_{0011}^2 + B_{0000}^2\right)
+2c_{23}B_{0000}^3  \,\,,
\end{align*}
\begin{align*}
D_{13}&=\!-\!b_{13} B_{001}^3\!-\!b_{21}B_{001}^1\!-\!2c_{13} \!\left( (p_3^2\!-\!p_1\.p_3 ) B_{0011}^3\!+\! B_{0000}^3 \right)
\!-\!2c_{21} \left(2 (p_1^2\! -\! p_1\.p_2) B_{0011}^1\! +\! B_{0000}^1\right)\\
&+2 c_{22} B_{0000}^2
-2c_{23} \left( (p_3^2 - p_2\.p_3) B_{0011}^1 + B_{0000}^3\right)
-2c_{31}\left(2 (p_1^2 - p_1\.p_3) B_{0011}^1 + B_{0000}^1\right)  \,\,,
\end{align*}
\begin{align*}
D_{22}&=a_2 B_{11}^2+b_{12} \left( 2 B_{001}^2 + (p_2^2-p_1\.p_2 ) B_{111}^2\right)
+b_{22}\left(2 B_{001}^2 + (p_2^2 - p_2\.p_3) B_{111}^2\right)
+2c_{11}B_{0000}^1\\
&
+c_{12}\left((5 p_2^2 - 6 p_1\.p_2 + p_1^2) B_{0011}^2 +
 2 B_{0000}^2 + (p_2^2 - p_1\.p_2)^2 B_{1111}^2\right)\\
&
+2c_{22}\left(( 5 p_2^2 -3( p_1\.p_2+p_2\.p_3) + p_1\.p_3 ) B_{0011}^2 +
 2 B_{0000}^2 + (p_2^2 - p_1\.p_2) (p_2^2 - p_2\.p_3) B_{1111}^2\right)\\
&
+c_{32}\left((5 p_2^2 - 6 p_2\.p_3 + p_3^2) B_{0011}^2 +
 2 B_{0000}^2 + (p_2^2 - p_2\.p_3)^2 B_{1111}^2\right)
+2 c_{33}B_{0000}^3  \,\,,
\end{align*}
\begin{align*}
D_{23}&=D_{32}=-b_{22} B_{001}^2-b_{23}B_{001}^3+2c_{21}B_{0000}^1
-2c_{22}\left((p_2^2 - p_1\.p_2) B_{0011}^2 + B_{0000}^2\right)\\
&-2c_{23}\left( (p_3^2 - p_1\.p_3) B_{0011}^3 + B_{0000}^3\right)
-2c_{32}\left((p_2^2 - p_2\.p_3) B_{0011}^2 + B_{0000}^2\right)\\
&-2c_{33}\left( (p_3^2 - p_2\.p_3) B_{0011}^3 + B_{0000}^3\right)  \,\,,
\end{align*}
\begin{align}\label{Dexpr}
D_{33}&=a_3 B_{11}^3+b_{13} \left(2 B_{001}^3 + (p_3^2-p_1\.p_3 ) B_{111}^3\right)
+b_{23}\left(2 B_{001}^3 + (p_3^2 - p_2\.p_3) B_{111}^3\right) \nonumber \\
&
+c_{13}\left((5 p_3^2- 6 p_1\.p_3+p_1^2  ) B_{0011}^3 +
 2 B_{0000}^3 + (p_3^2 - p_1\.p_3)^2 B_{1111}^3\right) \nonumber  \\
&
+2c_{23}\left((5 p_3^2- 3 (p_1\.p_3 + p_2\.p_3) + p_1\.p_2  ) B_{0011}^3 +
 2 B_{0000}^3 + (p_3^2 - p_1\.p_3) (p_3^2 - p_2\.p_3) B_{1111}^3\right) \nonumber \\
&
+2c_{31}B_{0000}^1
+2c_{32}B_{0000}^2
+c_{33}\left((5 p_3^2 - 6 p_2\.p_3 + p_2^2) B_{0011}^3 +
 2 B_{0000}^3 + (p_3^2 - p_2\.p_3)^2 B_{1111}^3\right)  \,\,,
\end{align}
with,
\begin{align*}
a_1&=\frac{\A_1}{(1+\chi_{12}) (1+\chi_{13})},\quad a_2=\frac{\A_2}{(1+\chi_{21}) (1+\chi_{23})},\quad
a_3=\frac{\A_3}{(1+\chi_{31}) (1+\chi_{32})}\\
b_{11}&=\frac{2 \A_1}{(1+\chi_{12})^2(1+\chi_{13}) (m_1^2-m_2^2)},\quad
b_{21}=\frac{2 \A_1}{(1+\chi_{12}) (1+\chi_{13})^2 (m_1^2-m_3^2)},\\
b_{12}&=\frac{2 \A_2}{(1+\chi_{21})^2 (1+\chi_{23}) (m_2^2-m_1^2)},\quad
b_{22}=\frac{2 \A_2}{(1+\chi_{21}) (1+\chi_{23})^2 (m_2^2-m_3^2)},\\
b_{13}&=\frac{2 \A_3}{(1+\chi_{31})^2(1+\chi_{32}) (m_3^2-m_1^2)},\quad
b_{23}=\frac{2 \A_3}{(1+\chi_{31}) (1+\chi_{32})^2(m_3^2-m_2^2)} ,\\
c_{11}&=\frac{4 \A_1}{(1+\chi_{12})^3 (1+\chi_{13}) (m_1^2-m_2^2)^2} ,\quad
c_{21}=\frac{2 \A_1}{(1+\chi_{12})^2 (1+\chi_{13})^2 (m_1^2-m_2^2) (m_1^2-m_3^2)} ,\\
c_{31}&=\frac{4 \A_1}{(1+\chi_{12}) (1+\chi_{13})^3 (m_1^2-m_3^2)^2},\quad
c_{12}=\frac{4 \A_2}{(1+\chi_{21}) (1+\chi_{21})^3 (m_2^2-m_1^2)^2}\\
c_{22}&=\frac{2 \A_2}{(1+\chi_{21})^2 (1+\chi_{23})^2(m_2^2-m_1^2) (m_2^2-m_3^2)},\quad
c_{32}=\frac{4 \A_2}{(1+\chi_{21})(1+\chi_{23})^3 (m_2^2-m_3^2)^2},\\
c_{13}&=\frac{4 \A_3}{(1+\chi_{31})^3 (1+\chi_{32}) (m_3^2-m_1^2)^2},\quad
c_{23}=\frac{2 \A_3}{(1+\chi_{31})^2 (1+\chi_{32})^2 (m_3^2-m_1^2) (m_3^2-m_2^2)} \\
c_{33}&=\frac{4 \A_3}{(1+\chi_{31}) (1+\chi_{32})^3(m_3^2-m_2^2)^2} \,\, .
\end{align*}
	
\bibliography{kpark}% Produces the bibliography via BibTeX.

\begin{thebibliography}{50}
\expandafter\ifx\csname natexlab\endcsname\relax\def\natexlab#1{#1}\fi
\expandafter\ifx\csname bibnamefont\endcsname\relax
  \def\bibnamefont#1{#1}\fi
\expandafter\ifx\csname bibfnamefont\endcsname\relax
  \def\bibfnamefont#1{#1}\fi
\expandafter\ifx\csname citenamefont\endcsname\relax
  \def\citenamefont#1{#1}\fi
\expandafter\ifx\csname url\endcsname\relax
  \def\url#1{\texttt{#1}}\fi
\expandafter\ifx\csname urlprefix\endcsname\relax\def\urlprefix{URL }\fi
\providecommand{\bibinfo}[2]{#2}
\providecommand{\eprint}[2][]{\url{#2}}

\bibitem[{\citenamefont{Denner and Dittmaier}(2006)}]{Denner:2005nn}
\bibinfo{author}{\bibfnamefont{A.}~\bibnamefont{Denner}} \bibnamefont{and}
  \bibinfo{author}{\bibfnamefont{S.}~\bibnamefont{Dittmaier}},
  \bibinfo{journal}{Nucl. Phys.} \textbf{\bibinfo{volume}{B734}},
  \bibinfo{pages}{62} (\bibinfo{year}{2006}), \eprint{hep-ph/0509141}.

\bibitem[{\citenamefont{Davydychev}(1991)}]{Davydychev:1991va}
\bibinfo{author}{\bibfnamefont{A.~I.} \bibnamefont{Davydychev}},
  \bibinfo{journal}{Phys. Lett.} \textbf{\bibinfo{volume}{B263}},
  \bibinfo{pages}{107} (\bibinfo{year}{1991}).

\bibitem[{\citenamefont{Denner}(1993)}]{Denner:1991kt}
\bibinfo{author}{\bibfnamefont{A.}~\bibnamefont{Denner}},
  \bibinfo{journal}{Fortschr. Phys.} \textbf{\bibinfo{volume}{41}},
  \bibinfo{pages}{307} (\bibinfo{year}{1993}), \eprint{0709.1075}.

\bibitem[{\citenamefont{Bern et~al.}(1994)\citenamefont{Bern, Dixon, and
  Kosower}}]{Bern:1993kr}
\bibinfo{author}{\bibfnamefont{Z.}~\bibnamefont{Bern}},
  \bibinfo{author}{\bibfnamefont{L.~J.} \bibnamefont{Dixon}}, \bibnamefont{and}
  \bibinfo{author}{\bibfnamefont{D.~A.} \bibnamefont{Kosower}},
  \bibinfo{journal}{Nucl. Phys.} \textbf{\bibinfo{volume}{B412}},
  \bibinfo{pages}{751} (\bibinfo{year}{1994}), \eprint{hep-ph/9306240}.

\bibitem[{\citenamefont{Binoth et~al.}(2000)\citenamefont{Binoth, Guillet, and
  Heinrich}}]{Binoth:1999sp}
\bibinfo{author}{\bibfnamefont{T.}~\bibnamefont{Binoth}},
  \bibinfo{author}{\bibfnamefont{J.~P.} \bibnamefont{Guillet}},
  \bibnamefont{and} \bibinfo{author}{\bibfnamefont{G.}~\bibnamefont{Heinrich}},
  \bibinfo{journal}{Nucl. Phys.} \textbf{\bibinfo{volume}{B572}},
  \bibinfo{pages}{361} (\bibinfo{year}{2000}), \eprint{hep-ph/9911342}.

\bibitem[{\citenamefont{Denner and Dittmaier}(2003)}]{Denner:2002ii}
\bibinfo{author}{\bibfnamefont{A.}~\bibnamefont{Denner}} \bibnamefont{and}
  \bibinfo{author}{\bibfnamefont{S.}~\bibnamefont{Dittmaier}},
  \bibinfo{journal}{Nucl. Phys.} \textbf{\bibinfo{volume}{B658}},
  \bibinfo{pages}{175} (\bibinfo{year}{2003}), \eprint{hep-ph/0212259}.

\bibitem[{\citenamefont{Passarino and Veltman}(1979)}]{Passarino:1978jh}
\bibinfo{author}{\bibfnamefont{G.}~\bibnamefont{Passarino}} \bibnamefont{and}
  \bibinfo{author}{\bibfnamefont{M.~J.~G.} \bibnamefont{Veltman}},
  \bibinfo{journal}{Nucl. Phys.} \textbf{\bibinfo{volume}{B160}},
  \bibinfo{pages}{151} (\bibinfo{year}{1979}).

\bibitem[{\citenamefont{Murley and Saad}(1979)}]{Murley:2008}
\bibinfo{author}{\bibfnamefont{J.}~\bibnamefont{Murley}} \bibnamefont{and}
  \bibinfo{author}{\bibfnamefont{N.}~\bibnamefont{Saad}},
 \eprint{math-ph/08095203}, (\bibinfo{year}{2008}).

\bibitem[{\citenamefont{Hahn}(1979)}]{Hahn:2000jm}
\bibinfo{author}{\bibfnamefont{T.} \bibnamefont{Hahn}},
  \bibinfo{journal}{Nucl. Phys. Suppl.} \textbf{\bibinfo{volume}{89}},
  \bibinfo{pages}{231} (\bibinfo{year}{2000}).

\bibitem[{\citenamefont{'tHooft}(1978)}]{'tHooft:1978xw}
\bibinfo{author}{\bibfnamefont{G.} \bibnamefont{'tHooft}} \bibnamefont{and}
\bibinfo{author}{\bibfnamefont{M. J. G.} \bibnamefont{Veltman}},
  \bibinfo{journal}{Nucl. Phys. B} \textbf{\bibinfo{volume}{153}},
  \bibinfo{pages}{365} (\bibinfo{year}{1979}).

\bibitem[{\citenamefont{Binosi}(1978)}]{Binosi:2008ig}
\bibinfo{author}{\bibfnamefont{D.} \bibnamefont{Binosi}},  
\bibinfo{author}{\bibfnamefont{J.} \bibnamefont{Collins}}, 
\bibinfo{author}{\bibfnamefont{C.} \bibnamefont{Kaufhold}}, \bibnamefont{and}
\bibinfo{author}{\bibfnamefont{L.} \bibnamefont{Theussl}},
  \bibinfo{journal}{Comput. Phys. Commun.} \textbf{\bibinfo{volume}{180}},
  \bibinfo{pages}{1709} (\bibinfo{year}{2009}).




\end{thebibliography}

\end{document}